\documentclass[a4paper,11pt]{article}

\usepackage{anysize}
\usepackage{amsthm}
\usepackage{amssymb}
\usepackage{amsmath}
\usepackage{fancybox}
\usepackage{amsfonts}
\usepackage[T1]{fontenc}
\usepackage{latexsym}
\usepackage{makeidx}
\usepackage{cite}
\usepackage[numbers]{natbib}

\usepackage[pdftex]{graphicx}
\usepackage[pdftex]{color}

\usepackage[all]{xy}

\newtheorem{teo}{Theorem}
\newtheorem{lem}{Lemma}
\newtheorem{pro}{Proposition}
\newtheorem{cor}{Corollary}
\newtheorem{defi}{Definition} 

\newcommand{\fd}{\rightarrow}

\newcommand{\inc}{\subset}

\newcommand{\ba}{\overline}

\newcommand{\al}{\alpha}
\newcommand{\be}{\beta}

\newcommand{\fhi}{\varphi}
\newcommand{\del}{\delta}
\newcommand{\Del}{\Delta}

\newcommand{\Gam}{\Gamma}

\newcommand{\Om}{\Omega}

\newcommand{\Z}{\mathbb{Z}}
\newcommand{\N}{\mathbb{N}}

\newcommand{\R}{\mathbb{R}}

\newcommand{\B}{\mathbb{B}}

\newcommand{\Sa}{\mathbb{S}}

\newcommand{\pa}{\partial}

\newcommand{\p}{\grave{}}

\newtheorem{remark}{Remark}[section]
\newtheorem{example}{Example}

\def\pf{\par\noindent {\em Proof.}~\par\noindent}

\def\lim{\mathop{\mbox{\normalfont lim}}\limits}

\def\pf{\par\noindent {\em Proof. }}

\def\pa{\partial}

\begin{document}

\date{}

\title{Distributions and Integration in superspace}
\small{
\author
{Al\'i Guzm\'an Ad\'an, Frank Sommen}
\vskip 1truecm
\date{\small  Clifford Research Group, Department of Mathematical Analysis, Faculty of Engineering
and Architecture, Ghent University, Krijgslaan 281, 9000 Gent, Belgium. \\
{\tt Ali.GuzmanAdan@UGent.be}, \hspace{.3cm}{\tt Franciscus.Sommen@UGent.be} }

\maketitle

\begin{abstract} Distributions in superspace constitute a very useful tool for establishing an integration theory. In particular, distributions have been used to obtain a suitable extension of the Cauchy formula to superspace and to define integration over the superball and the supersphere through the Heaviside and Dirac distributions, respectively.

In this paper, we extend the distributional approach to integration over more general domains and surfaces in superspace. The notions of domain and surface in superspace are defined by smooth bosonic phase functions $g$. This allows to define domain integrals and oriented (as well as non-oriented) surface integrals in terms of the Heaviside and Dirac distributions of the superfunction $g$. It will be shown that the presented definition for the integrals does not depend on the choice of the phase function $g$ defining the corresponding domain or surface. In addition, some examples of integration over a super-paraboloid and a super-hyperboloid will be presented. Finally, a new distributional Cauchy-Pompeiu formula will be obtained, which generalizes and unifies the previously known approaches.

\noindent

\vspace{0.3cm}

\small{ }

\noindent
\textbf{}
\end{abstract}

\section{Introduction}
Supermanifolds and in particular superspaces play an important r\^ole in contemporary theoretical physics, e.g.\ in the particle theory of supersymmetry, supergravity or superstring theories, etc. Superspaces are equipped with both a set of commuting co-ordinates and a set of anti-commuting co-ordinates. From the mathematical point of view, they have been studied using algebraic and geometrical methods (see e.g.\ \cite{Berezin:1987:ISA:38130, MR565567, MR778559}). More recently, harmonic and Clifford analysis have been extended to superspace by introducing some important differential operators (such as Dirac and Laplace operators) and by studying special functions and orthogonal polynomials related to these operators, see e.g.\ \cite{Bie2007, de2007clifford, MR2344451, 1751-8121-42-24-245204, MR2521367, MR2683546}. 

Integration on superspace is based on the notion of the Berezin integral given by $\int_B=\pi^{-n} \, \pa_{x\p_{2n}}\cdots \pa_{x\p_{1}}$, see \cite{Berezin:1987:ISA:38130}. This functional plays the same r\^ole in the Grassmann algebra generated by the anti-commuting variables $x\p_1,\ldots, x\p_{2n}$ as the general real integral $\int_{\R^m} dV_{\underline x}$ in classical analysis. Traditionally, the Berezin integral is combined with the classical real integration in order to integrate superfunctions over real domains, i.e.\ the integral of a superfunction $F$ over $\Om\inc\R^m$ is given by
\begin{equation}\label{RealIntSupSpa}
\int_{\Om} \int_B F(\underline{x}, \underline{x}\p) \, dV_{\underline x}. 
\end{equation}
Some important classical results, such as a Stokes and a Cauchy-Pompeiu formula have been {established} for the super Dirac operator, see  \cite{MR2521367}. {Yet} these extensions have important limitations since they only consist of real integration combined with the Berezin integral, {insted of considering} general integration over domains and surfaces defined in terms of both commuting and anti-commuting co-ordinates in superspace. 

The study of spherical harmonics (and monogenics) has lead to an important development of integration theory in superspace. For example, in \cite{MR2344451, 1751-8121-42-24-245204} the Berezin integral was related to more familiar types of integration like Pizzetti's formula, see \cite{Pizz}. In this way, the integral of a polynomial $P$ over the supersphere was introduced as:
\begin{equation}\label{PizzSS}
\int_{SS} P=\sum_{j=0}^\infty(-1)^j \frac{2\pi^{M/2}}{2^{2j}\, j!\, \Gamma(j+M/2)} (\Del^j P)(0),
\end{equation}
where $\Del$ is the super Laplace operator and $M$ the corresponding superdimension.  In the later work \cite{MR2539324}, formula (\ref{PizzSS}) was extended to more general superfunctions on the supersphere by considering the integral
\begin{equation}\label{SSIntSupSpa}
\int_{SS} F=2 \int_{\R^m} \int_B \del({\bf x}^2+1) F \,dV_{\underline x},
\end{equation}
where $\del({\bf x}^2+1)$ denotes the Dirac distribution on the unit supersphere. 
Following this last distributional approach, some important problems were solved. In particular, closed formulas for the Pizzetti integral, the Funk-Hecke theorem and a Cauchy-Pompeiu formula  for the supersphere were obtained, see \cite{MR2539324, MR2683546}.

Nevertheless, the approaches given in (\ref{RealIntSupSpa}) and (\ref{SSIntSupSpa}) are still limited. They only refer to the particular cases of integration of superfunctions over real domains or over the supersphere. The main goal of this paper is to extend and unify both approaches by defining integration over general domains and surfaces in superspace. The idea of this extension is inspired by H\"ormander's formulas, which provide a distributional approach to classical real integration, see \cite[Theorem 6.1.5]{MR1996773}.
Indeed, suppose that $\Om\inc \R^m$ is a domain ($m$ dimensional manifold) determined by some inequality $g_0(x_1, \ldots, x_m) <0$  and let $\pa\Om$ be its boundary ($m-1$ dimensional manifold in $\R^m$) determined by the equation $g_0(x_1, \ldots, x_m) =0$. Then the integrals over $\Om$ and $\pa\Om$ can be rewritten as
\begin{equation}
\int_\Om (\cdot) \,dV_{\underline{x}}= \int_{\R^m} H(-g_0(\underline{x})) (\cdot)\, dV_{\underline{x}}, \;\;\mbox{ and } \;\;\int_{\pa\Om} \, (\cdot)\,dS_{\underline{w}}=\int_{\R^m} \del(g_0(\underline{x}))\, |\pa_{\underline{x}}[g_0](\underline{x})|\, (\cdot)\; dV_{\underline{x}},
\end{equation}
respectively; with $\underline{x}=(x_1, \ldots, x_m)$, $H$ the Heaviside distribution and $\del$ the Dirac distribution. In this way, one may see the integrals $\int_\Om$ and $\int_{\pa\Om}$ not as functionals  depending on geometrical sets of points $\Om$ and $\pa\Om$; but as functionals depending on the action of the Heaviside or Dirac distributions on a fixed phase function $g_0$.

 As will be shown in this paper, this last approach is the more suitable one to extend domain and surface integrals to superspace. In particular, we will illustrate it by integrating over a super-paraboloid and super-hyperboloid.  Moreover, this approach will be used to obtain a Cauchy-Pompeiu formula, valid not only for real domains and  for the superball, but for every domain with smooth boundary in superspace. 
 This allows to follow a completely analytical method which uses the Cauchy kernel as a true distribution rather than as a smooth function with a singularity at the origin. This distributional Cauchy-Pompeiu formula will play an essential r\^ole {in obtaining} a Bochner-Martinelli formula for holomorphic functions in superspace. That will be the topic of future work. 

The paper is organized as follows. In section 2, we provide some basics on harmonic and Clifford analysis in superspace. In section 3, we give a brief overview on the general theory of distributions in superspace. We pay particular attention to the Dirac and Heaviside distributions  of bosonic smooth  phase functions. 
In section 4, the definition of domain and surface integrals in superspace are formally given following the above mentioned distributional approach.  We prove that these definitions do not depend on the chosen superfunction defining the corresponding domain or surface. Section 5 is devoted to some examples of integration over a super-paraboloid and a super-hyperboloid of revolution. In particular we compute the volumes and surface areas produced by these objects when considering a positive height $h>0$. This leads to very interesting extensions of the classical formulas in 2 and 3 real dimensions, in terms of hypergeometric functions depending on the superdimension $M$. Finally, in section 6 we prove the distributional Cauchy-Pompeiu formula which extends and unifies the already known results in \cite{MR2521367, MR2539324}.

\section{Preliminaries}

Superanlaysis or analysis on superspace considers not only commuting (bosonic) but also anticommuting (fermionic) variables. In this paper we follow the extension of harmonic and Clifford analysis to superspace (see \cite{de2007clifford, Bie2007, MR2344451, 1751-8121-42-24-245204}). This approach considers $m$ commuting variables $x_1,\ldots, x_m$ and $2n$ anti-commuting variables $x\p_1, \ldots, x\p_{2n}$ in a purely symbolic way, i.e.\
\begin{align*}
 x_jx_k &= x_kx_j, & x\p_j x\p_k &= -x\p_k x\p_j, & x_jx\p_k &= x\p_k x_j. 
\end{align*}
This means that $x_1,\ldots, x_m$ are interpreted as generators of the polynomial algebra $\R[x_1,  \ldots, x_m]$ while $x\p_1, \ldots, x\p_{2n}$ generate a Grassmann algebra $\mathfrak{G}_{2n}$. We will denote by $\mathfrak{G}^{(ev)}_{2n}$ and $\mathfrak{G}^{(odd)}_{2n}$ the subalgebras of even and odd elements of $\mathfrak{G}_{2n}$ respectively. The algebra of super-polynomials, i.e. polynomials in the variables $x_1,  \ldots, x_m, x\p_1, \ldots, x\p_{2n}$ with coefficients in $\R$, is defined by
\[\mathcal{P}:= \mbox{Alg}_{\R}(x_1,  \ldots, x_m, x\p_1, \ldots, x\p_{2n})= \R[x_1,  \ldots, x_m] \otimes \mathfrak{G}_{2n}.\]

The bosonic and fermionic partial derivatives $\pa_{x_j}$, $\pa_{x\p_j}$ are defined purely symbolically as endomorphisms on $\mathcal P$ by the  relations 
\begin{equation*}
\begin{cases} \pa_{x_j}[1]=0,\\
 \pa_{x_j} x_k- x_k \pa_{x_j}=\del_{j,k},\\
 \pa_{x_j} x\p_k=x\p_k \pa_{x_j}, \;\; 
 \end{cases}
\hspace{.5cm} 
\begin{cases} \pa_{x\p_j}[1]=0,\\
\pa_{x\p_j} x\p_k+ x\p_k \pa_{x\p_j}=\del_{j,k}\\
\pa_{x\p_j} x_k=x_k\pa_{x\p_j}, 
\end{cases}
\end{equation*} 
that can be applied recursively. From this definition it immediately follows that $\pa_{x_j} \pa_{x_k}=\pa_{x_k}\pa_{x_j}$, $ \pa_{x\p_j}\pa_{x\p_k}=-\pa_{x\p_k}\pa_{x\p_j}$ and $\pa_{x_j}\pa_{x\p_k}=\pa_{x\p_k}\pa_{x_j}$.

Clifford algebras in superspace are introduced by $m$ orthogonal Clifford generators  $e_1,\ldots, e_m$ and $2n$ symplectic Clifford generators $e\p_1, \ldots, e\p_{2n}$ which are subjected to the multiplication relations
\[e_je_k+e_ke_j=-2\del_{j,k}, \hspace{.3cm} e_je\p_k+e\p_ke_j=0, \hspace{.3cm} e\p_je\p_k-e\p_ke\p_j=g_{j,k},\]
where $g_{j,k}$ is a symplectic form defined by
\[g_{2j,2k}=g_{2j-1,2k-1}=0, \hspace{.5cm} g_{2j-1,2k}=-g_{2k,2j-1}=\del_{j,k}, \hspace{.5cm} j,k=1,\ldots,n.\]
These generators are combined with the algebra of super-polynomials $\mathcal{P}$  giving rise to the algebra of Clifford valued super-polynomials
\[\mathcal{P} \otimes\mathcal C_{m,2n} =\mbox{Alg}_\R(x_1,\ldots, x_m, x\p_1, \ldots, x\p_{2n}, e_1,\ldots,e_m,e\p_1,\ldots,e\p_{2n})\]
where elements in $\mathcal C_{m,2n}=\mbox{Alg}_\R(e_1,\ldots,e_m,e\p_1,\ldots,e\p_{2n})$ commute  with elements in $\mathcal{P}$. Also the partial derivatives $\pa_{x_j}$, $\pa_{x\p_j}$ commute with the elements in the algebra $\mathcal C_{m,2n}$. The most important element of the algebra  $\mathcal{P} \otimes\mathcal C_{m,2n}$ is the supervector variable
\begin{equation*}
{\bf x}=\underline{x}+\underline{x\p}=\sum_{j=1}^mx_je_j+\sum_{j=1}^{2n}x\p_j e\p_j.
\end{equation*}

Analysis in superspace studies functions of a supervector variable ${\bf x}=\underline{x}+\underline{x\p}\,$ of the form
\begin{equation}\label{SupFunc}
F({\bf x})=F(\underline{x},\underline{x\p})=\sum_{A\inc \{1, \ldots, 2n\}} \, F_A(\underline{x})\, \underline{x}\p_A, 
\end{equation}
where $ \underline{x}\p_A$ is defined as $x\p_{j_1}\ldots x\p_{j_k}$ with $A=\{j_1,\ldots, j_k\}$ $(1\leq j_1< \ldots< j_k\leq 2n)$, and $F_A(\underline{x})$  is a  function depending only on the bosonic variables $x_1,\ldots, x_m$. The most basic set of superfunctions is the super-polynomial algebra $\mathcal{P}$.  We can consider more general function spaces based on the properties of the  functions $F_A$. 
Indeed, given a bosonic function space $\mathcal F=C^k(\R^m), L_2(\Om), \ldots$, we obtain  a corresponding space of superfunctions $\mathcal F\otimes \mathfrak{G}_{2n}$.

The  bosonic and fermionic Dirac operators are defined 
by
\[\pa_{\underline x}=\sum_{j=1}^m e_j\pa_{x_j}, \hspace{1cm} \pa_{\underline x\p}=2\sum_{j=1}^n \left(e\p_{2j}\pa_{x\p_{2j-1}}-e\p_{2j-1}\pa_{x\p_{2j}}\right),\]
which lead to the left and right super Dirac operators (super-gradient)
$\pa_{\bf x} \cdot =\pa_{\underline x\p}\cdot -\pa_{\underline x}\cdot$ and $\cdot\pa_{\bf x}  =- \cdot\pa_{\underline x\p} - \cdot \pa_{\underline x}$, respectively. As in the classical Clifford setting, the action of $\pa_{\bf x}$ on the vector variable ${\bf x}$ results in the superdimension $M=m-2n$. {In this paper we work with general superdimensions $M$ (but $m\neq 0$).

Associated to the usual gradation of the Grassmann algebra $\mathfrak{G}_{2n}$, we consider for every $F$ of the form (\ref{SupFunc}) the superfunction $F^*({\bf x})=\sum_{A\inc\{1,\ldots, 2n\}} (-1)^{|A|} F_A(\underline{x})  \underline{x}\p_A$.
It can be easily proved that 
\begin{align}\label{Hash}
(F^*)^*&=F,   &     \pa_{x\p_j}[F]&=-[F^*]\pa_{x\p_j},   &  \pa_{x\p_j}[FG]&=\pa_{x\p_j}[F]G+ F^*\pa_{x\p_j}[G].
\end{align}

Following the classical approach, the (real) support $supp\; F$ of a superfunction $F\in C^\infty(\R^m) \otimes \mathfrak{G}_{2n}$ is defined as the closure of the set of all points in $\R^m$ for which the function $F(\cdot, \underline{x}\p):\R^m\fd \mathfrak{G}_{2n}$ is not zero, see (\ref{SupFunc}). From this definition, it immediately follows  that $supp\; F =\bigcup_{A\inc \{1, \ldots, 2n\}} supp\; F_A$.




Every superfunction can be written as the sum $F({\bf x})= F_0(\underline{x})+ {\bf F}(\underline x, \underline{x}\p)$ where the real valued function $F_0(\underline{x})=F_\emptyset(\underline{x})$ is called the {\it body} $F$, and ${\bf F}=\sum_{|A|\geq 1} \, F_A(\underline{x})\,\underline{x}\p_A$ is the {\it nilpotent part} of $F$. Indeed, it is clearly seen that  ${\bf F}^{2n+1}=0$.


 
 It is possible to produce interesting even superfunctions out of known functions from real analysis. Indeed, 
consider a smooth function $F\in C^\infty(\R)$ and an even superfunction $a=a_0+{\bf a}\in C^\infty(\R^m)\otimes \mathfrak{G}^{(ev)}_{2n}$ where $a_0$ and ${\bf a}$ are the body and nilpotent part of $a$, respectively. Then the superfunction $F(a({\bf x}))\in C^\infty(\R^m)\otimes \mathfrak{G}^{(ev)}_{2n}$ is defined, through the Taylor expansion of $F$, as
\begin{equation}\label{BosSuFun}
F(a)=F(a_0+{\bf a})=\sum_{j=0}^{n} \;\frac{{\bf a}^j}{j!}\;F^{(j)}(a_0).
\end{equation}
Straightforward calculations show that the  above expression is independent of the splitting of the even superfunction $a$ if the function $F$ is analytic in $\R$.
\begin{pro}\label{SplitAnalFunc}
Let $a,b\in  C^\infty(\R^m)\otimes \mathfrak{G}_{2n}^{(ev)}$ be such that $a=a_0+{\bf a}$, $b=b_0+{\bf b}$, where $a_0,b_0$ are the bodies of $a$, $b$ respectively and ${\bf a}$, ${\bf b}$ are the corresponding nilpotent projections. Then, for every analytic function $F\in C^\infty(\R)$ the following statements hold.
\begin{itemize}
\item[i)] $F(a+b)=\sum_{j=0}^{n} \; \frac{{\bf b}^j}{j!}\; F^{(j)}(a+b_0)$,
\item[ii)] $F(a+b)=\sum_{j=0}^{\infty} \; \frac{b_0^j}{j!}\; F^{(j)}(a+{\bf b})$,
\item[iii)] $F(a+b)=\sum_{j=0}^{\infty} \; \frac{b^j}{j!}\; F^{(j)}(a)$.
\end{itemize}
\end{pro}


The easiest application of the extension (\ref{BosSuFun}) is obtained when defining arbitrary real powers of even superfunctions.  Let $p\in \R$ and $a=a_0+{\bf a}\in C^\infty(\R^m)\otimes \mathfrak{G}_{2n}^{(ev)}$, then for $a_0>0$ we define  
\[a^p=\sum_{j=0}^{n}\;\frac{{\bf a}^j}{j!}\; (-1)^j\,(-p)_j \,a_0^{p-j}, \hspace{.5cm} \mbox{ where }  \hspace{.5cm} (q)_j=\begin{cases} 1, & j=0,\\ q(q+1)\cdots (q+j-1), & j>0, \end{cases}\]
is the rising Pochhammer symbol. Observe that {if the numbers $q$ and $q+j$ are in the set $\R\setminus\{0, -1, -2,\ldots\}$} we can write $\displaystyle (q)_j=\frac{\Gam(q+j)}{\Gam(q)}$. Making use of this definition of power function in superspace, we can easily prove that its basic properties still hold in this setting.



\begin{lem}\label{Pow}
Let $a=a_0+{\bf a}$ and  $b=b_0+{\bf b}$ be elements in $C^\infty(\R^m)\otimes \mathfrak{G}_{2n}^{(ev)}$ such that $a_0,b_0>0$. Then, for every pair $p,q\in \R$ we have
\[i) \;a^pa^q=a^{p+q}, \hspace{1cm} ii)\; (ab)^p=a^pb^p, \hspace{1cm} iii) \;(a^p)^q=a^{pq}.\]
\end{lem}
\pf
For $a_0,b_0>0$, the equalities
\[(a_0+X)^p(a_0+X)^q=(a_0+X)^{p+q}, \hspace{.25cm} \left[(a_0+X)(b_0+Y)\right]^p=(a_0+X)^p(b_0+Y)^p, \hspace{.25cm} \left((a_0+X)^p\right)^q=(a_0+X)^{pq},\]
are identities in formal power series in the indeterminates $X$ and $Y$. Then, making the substitutions $X={\bf a}$ and $Y={\bf b}$ we obtain $i)$, $ii)$ and $iii)$. Observe that the nilpotency of ${\bf a}$ and ${\bf b}$ avoids every possible convergence issue. $\hfill\square$

The absolute value function can be defined over $C^\infty(\R^m)\otimes \mathfrak{G}_{2n}^{(ev)}$ by
\[|a|=(a^2)^{1/2}=\begin{cases} a & \mbox{ if }\;\; a_0\geq 0, \\ -a &  \mbox{ if } \;\;a_0< 0. \end{cases}\]
As in the classical case, the absolute value function can be extended to the supervector variable ${\bf x}$, since its square is an even super-polynomial, i.e.
\[{\bf x}^2=-\sum_{j=1}^m x_j^2 + \sum_{j=1}^n x\p_{2j-1} x\p_{2j}\in C^\infty(\R^m)\otimes \mathfrak{G}_{2n}^{(ev)}.\]
It is clear that $-{\bf x}^2$ has non-negative body. Hence, the element $(-{\bf x}^2)^{1/2}$ is well defined. In this way, we define the absolute value of a supervector by
\[|{\bf x}|=(-{\bf x}^2)^{1/2}=\left(|\underline{x}|^2- \underline{x\p}^2 \right)^{1/2}=\sum_{j=0}^n \frac{(-1)^j \underline{x\p}^{\,2j}}{j!} \, \frac{\Gam(\frac{3}{2})}{\Gam(\frac{3}{2}-j)} |\underline{x}|^{1-2j}, \;\;\mbox{ where } \;\; |\underline{x}|=\left(\sum_{j=1}^m x_j^2\right)^{\frac{1}{2}}.\]

\begin{pro}\label{PropMod}
Let ${\bf x},{\bf y}$ be supervector variables and $a\in C^\infty(\R^m)\otimes \mathfrak{G}_{2n}^{(ev)}$. Then,
\begin{itemize}
\item[i)] $|a{\bf x}|=|a||{\bf x}|,$
\item[ii)] $|{\bf x}+a{\bf y}|=|{\bf x}|+aF({\bf x},{\bf y},a)$ where $F({\bf x},{\bf y},a)$ is a superfunction depending on ${\bf x}, {\bf y}$ and $a$.
\end{itemize}
\end{pro}
\pf
\begin{itemize}
\item[$i)$] By Lemma \ref{Pow} we get $|a{\bf x}|=\left(-(a{\bf x})^2\right)^{\frac{1}{2}}=\left(a^2(-{\bf x}^2)\right)^{\frac{1}{2}}=\left(a^2\right)^{\frac{1}{2}}\left(-{\bf x}^2\right)^{\frac{1}{2}}=|a||{\bf x}|$.
\item[$ii)$] We first write $({\bf x}+a{\bf y})^2={\bf x}^2+a\{{\bf x},{\bf y}\}+a^2{\bf y}^2={\bf x}^2-av$, where $v=-\{{\bf x},{\bf y}\}-a {\bf y}^2$ is an even element. Then, using Proposition \ref{SplitAnalFunc} $iii)$, we get
\[|{\bf x}+a{\bf y}| = \left(-({\bf x}+a{\bf y})^2\right)^\frac{1}{2}=\left(-{\bf x}^2+av\right)^\frac{1}{2}= \sum_{j=0}^\infty \frac{(a v)^j}{j!} \frac{\Gam(\frac{3}{2})}{\Gam(\frac{3}{2}-j)} (-{\bf x}^2)^{\frac{1}{2}-j}=|{\bf x}|+a F({\bf x},{\bf y},a), 
\]
where $ F({\bf x},{\bf y},a)= \sum_{j=1}^\infty \frac{a^{j-1} v^j}{j!} \frac{\Gam(\frac{3}{2})}{\Gam(\frac{3}{2}-j)} (-{\bf x}^2)^{\frac{1}{2}-j}$.
$\hfill\square$
\end{itemize}

\section{Distributions in superspace}
In this section we introduce the notion of distribution in superspace. In particular, we will study the extensions of the Heaviside and Dirac distributions to superspace. They play an important r\^ole in the definition of domain and surface integrals in superanalysis, as it will be shown in the next section. 

\subsection{Superdistributions}
Let $\mathcal{D}^\prime$ be the space of Schwartz distributions,  i.e. the space of generalized functions on the space $C^\infty_0(\R^m)$ of real valued $C^\infty$-functions with compact support. As usual, the notation
\begin{equation}\label{DistEv}
\int_{\R^m} \al f \; dV_{\underline x}= \langle\al,f\rangle,
\end{equation}
where $dV_{\underline{x}}=d{x_1}\cdots d{x_m}$ is the classical $m$-volume element, is used for the evaluation of the distribution $\al\in \mathcal{D}^\prime$ on the test function $f\in C^\infty_0(\R^m)$.

Let  $\mathcal{E}^\prime$ be the space of  generalized functions on the space  $C^\infty (\R^m)$ of $C^\infty$-functions in $\R^m$ (with arbitrary support). We recall that $\mathcal{E}^\prime$ is exactly the subspace of all compactly supported distributions in $\mathcal{D}^\prime$. Indeed, every distribution in $\mathcal{E}^\prime\inc \mathcal{D}^\prime$ has compact support and viceversa, every distribution in $\mathcal{D}^\prime$ with compact support can be uniquely extended to a distribution in $\mathcal{E}^\prime$, see \cite{MR0208364} for more details. This means that, for every $\al\in \mathcal{E}^\prime$, evaluations of the form (\ref{DistEv}) extend to $C^\infty (\R^m)$ (instead of $C^\infty_0 (\R^m)$).

The space of superdistributions  $\mathcal{D}^\prime \otimes \mathfrak{G}_{2n}$ {then} is defined by all elements of the form
\begin{equation}\label{SupDistForm}
\al=\sum_{A\inc\{1,\ldots, 2n\}} \al_A \underline{x}\p_A, \hspace{.5cm} \mbox{ with } \al_A\in \mathcal{D}^\prime.
\end{equation}
Similarly, the subspace $\mathcal{E}^\prime \otimes \mathfrak{G}_{2n}$ is composed by all elements of the form (\ref{SupDistForm}) but with $\al_A\in \mathcal{E}^\prime$.

The {analogue} of the integral $\int_{\R^m}\; dV_{\underline x}$ in superspace  is given by
\[\int_{R^{m|2n}}=\int_{\R^m} dV_{\underline{x}} \int_B=\int_B \int_{\R^m} dV_{\underline{x}},\]
where the bosonic integration is the usual real integration and the integral over fermionic variables is given  by the so-called Berezin integral (see \cite{Berezin:1987:ISA:38130}), defined by
\[\int_B=\pi^{-n} \, \pa_{x\p_{2n}}\cdots \pa_{x\p_{1}}=\frac{(-1)^n \pi^{-n}}{4^n n!} \pa_{\underline{x}\p}^{2n}.\]
This {enables us} to define the action of a superdistribution $\al\in \mathcal{D}^\prime \otimes \mathfrak{G}_{2n}$ (resp. $\al\in \mathcal{E}^\prime \otimes \mathfrak{G}_{2n}$) on a test superfunction $F\in C^\infty_0(\R^m) \otimes \mathfrak{G}_{2n}$ (resp. $F\in C^\infty(\R^m) \otimes \mathfrak{G}_{2n}$) by
\[\int_{\R^{m|2n}} \al F := \sum_{A,B\inc\{1,\ldots, 2n\}} \langle \al_A, f_B \rangle \int_B   \underline{x}\p_A\, \underline{x}\p_B. \]

As in the classical case, we say that the superdistribution $\al\in \mathcal{D}^\prime \otimes \mathfrak{G}_{2n}$ {\it vanishes in the open set} $\Om\inc \R^m$ if $\int_{\R^{m|2n}} \al F=0$ for every $F\in C^\infty_0(\R^m) \otimes \mathfrak{G}_{2n}$ whose real support is contained in $\Om$. In the same way, the support $supp\; \al$ of $\al\in \mathcal{D}^\prime \otimes \mathfrak{G}_{2n}$ is defined as the complement of the largest open subset of $\R^m$ on which $\al$ vanishes. Hence, it can be easily seen that $supp\; \al=\bigcup_{A\inc\{1,\ldots, 2n\}} supp\; \al_A$. This means that  $\mathcal{E}^\prime \otimes \mathfrak{G}_{2n}$ is the subspace of all compactly supported superdistributions in $\mathcal{D}^\prime \otimes \mathfrak{G}_{2n}$.

\subsection{Properties of the $\del$-distribution in real calculus.}
We now list some important properties of the $\del$-distribution in real calculus. This is necessary to introduce and study the main properties of the Heaviside distribution and all its derivatives in superspace.

%
\begin{pro}\label{1del}
Let $j,k\in\N\cup\{0\}$. Then $\del^{(j)}(x) \; x^k =\begin{cases} 0, & j<k,\\ (-1)^k\, k! \,\binom{j}{k} \, \del^{(j-k)}(x), & k\leq j.  \end{cases}$
\end{pro}

The proof of the above result is omitted since it runs along straightforward computations. In order to study the composition of the real $\del$-distribution with real valued functions in $\R^m$, we first need the following result.
\begin{pro}\label{DiracOpDist}
Let $g_0\in C^\infty(\R^m)$ such that $\pa_{\underline x}[g_0]\neq 0$ on the surface $g_0^{-1}(0):=\{\underline{w}\in\R^{m}: g_0(\underline{w})=0\}$. Then for every $j\in\N\cup\{0\}$, {it holds that}
\[\pa_{\underline{x}}\left[\del^{(j)}(g_0(\underline{x}))\right]=\pa_{\underline{x}}[g_0](\underline{x}) \, \del^{(j+1)}(g_0(\underline{x})).\]
\end{pro}
\pf
{From} the chain rule for partial derivatives acting on the composition $\del^{(j)}(g_0(\underline x))$ {it immediately follows that}
\begin{equation}\label{DervDelt}
\pa_{x_j}[\del(g_0^{(j)}(\underline{x}))]=\del^{(j+1)}(g_0(\underline{x})) \pa_{x_j}[g_0](\underline{x}),
\end{equation}
 see (6.1.2) in \cite[p.~135]{MR1996773}.
$\hfill\square$

\begin{pro}\label{ProMultH}
Let $g_0,h_0\in C^{\infty}(\R^m)$ be functions such that $h_0>0$ and $\pa_{\underline x}[g_0]\neq 0$ on the surface $g_0^{-1}(0)$. Then, for $j\in \N\cup\{0\}$ it holds that
\begin{equation}\label{MultH}
\del^{(j)}\left(h_0(\underline{x})g_0(\underline{x})\right)=\frac{\del^{(j)}(g_0(\underline{x}))}{h_0(\underline{x})^{j+1}}.
\end{equation}
\end{pro}
\pf
We proceed by induction on $j\in \N\cup\{0\}$. {In order} to prove the statement for $j=0$ we first observe that the following simple layer integral identity holds (see Theorem 6.1.5 in \cite{MR1996773}):
\begin{equation}\label{LayInt}
\int_{\R^m} \del(g_0(\underline{x})) f(\underline{x}) \; dV_{\underline{x}}=\int_{g_0^{-1}(0)} \frac{f(\underline{w})}{|\pa_{\underline{x}}[g_0](\underline{w})|}\; dS_{\underline{w}},
\end{equation}
where $dV_{\underline{x}}=d{x_1}\cdots d{x_m}$ is the classical $m$-{dimensional} volume element and $dS_{\underline{w}}$ is the Lebesgue surface measure on the surface $g_0^{-1}(0)$. We also have  $\pa_{\underline{x}}[h_0g_0]=\pa_{\underline{x}}[h_0]g_0+h_0\pa_{\underline{x}}[g_0]$, which implies $\pa_{\underline{x}}[h_0g_0](\underline{w})=h_0(\underline{w})\pa_{\underline{x}}[g_0](\underline{w})$ if $g_0(\underline{w})=0$.  Thus applying (\ref{LayInt}) to $g_0h_0$, instead of $g_0$, we get
\[\int_{\R^m} \del(h_0(\underline{x})g_0(\underline{x})) f(\underline{x}) \; dV_{\underline{x}}=\int_{g_0^{-1}(0)} \frac{f(\underline{w})}{h_0(\underline{w})\,|\pa_{\underline{x}}[g_0](\underline{w})|}\; dS_{\underline{w}}= \int_{\R^m} \frac{\del(g_0(\underline{x}))}{h_0(\underline{x})} f(\underline{x}) \; dV_{\underline{x}}, \]
for every real valued test function $f$.  Hence, $\del(h_0(\underline{x})g_0(\underline{x}))= \frac{\del(g_0(\underline{x}))}{h_0(\underline{x})}$.

\noindent {Now} assume (\ref{MultH}) to be true for $j\geq 1$. {We then} prove it for $j+1$. Letting $\pa_{\underline{x}}$ act on both sides of (\ref{MultH}) we get
\begin{equation}\label{Hypo}
\pa_{\underline{x}}[h_0]g_0\,\del^{(j+1)}(h_0g_0)+h_0\pa_{\underline{x}}[g_0]\,\del^{(j+1)}(h_0g_0)= \frac{\pa_{\underline{x}}[g_0]\,\del^{(j+1)}(g_0)}{h_0^{j+1}} \, - \, \frac{j+1}{h_0^{j+2}}\, \pa_{\underline{x}}[h_0] \, \del^{(j)}(g_0).
\end{equation}
By Proposition \ref{1del} and the induction hypothesis we get 
\[h_0g_0 \,\del^{(j+1)}(h_0g_0) = -(j+1) \del^{(j)}(h_0g_0)=-(j+1)\frac{\del^{(j)}(g_0)}{h_0^{j+1}}, \hspace{.2cm}\]
implying
  \[g_0 \,\del^{(j+1)}(h_0g_0)= -(j+1)\frac{\del^{(j)}(g_0)}{h_0^{j+2}}. \]
Substituting this in (\ref{Hypo}) we easily obtain
\[h_0\pa_{\underline{x}}[g_0]\,\del^{(j+1)}(h_0g_0)= \frac{\pa_{\underline{x}}[g_0]\,\del^{(j+1)}(g_0)}{h_0^{j+1}}  \hspace{.2cm} \mbox{ which is equivalent to } \hspace{.2cm}  \,\del^{(j+1)}(h_0g_0)= \frac{\,\del^{(j+1)}(g_0)}{h_0^{j+2}}.\]
Observe that the factor $\pa_{\underline{x}}[g_0]$ can be cancelled since $\pa_{x}[g_0]\neq 0$ in $g_0^{-1}(0)$. $\hfill\square$

\subsection{$\del$-Distribution in superspace}
In this section we introduce the $\del$-distribution in superspace together with all its derivatives. As usual, the Heaviside distribution will be introduced as the corresponding anti-derivative of the Dirac distribution. In  \cite{MR2539324}, these distributions were introduced for some particular cases corresponding to the supersphere.

From now on, we use the notation $p\N+q:=\{pk+q:k\in\N\}$ where $p,q\in \Z$ and $\N:=\{1,2,\ldots\}$ is the set of natural numbers.

Consider an even superfunction $g=g_0+{\bf g}\in C^\infty(\R^m)\otimes \mathfrak{G}_{2n}^{(ev)}$ such that $\pa_{\underline x}[g_0]\neq0$ on the surface $g_0^{-1}(0)$. The distribution $\del^{(k)}(g)$ is defined as the Taylor series  
\[\del^{(k)}(g) = \sum_{j=0}^{n} \frac{{\bf g}^j}{j!} \, \del^{(k+j)}(g_0), \hspace{1cm} k\in\N-2:=\{-1, 0, 1, 2\ldots\}.\]
The particular case $k=-1$ provides the expression for the antiderivative of $\del$, i.e. the Heaviside distribution $H=\del^{(-1)}$ given by
\begin{equation}\label{HaevSupS}
H(g)=H(g_0)+\sum_{j=1}^{n} \frac{{\bf g}^j}{j!} \, \del^{(j-1)}(g_0), \hspace{1cm} \mbox{ where } \hspace{.5cm}H(g_0)=\begin{cases}  1, & g_0\geq 0,\\ 0, & g_0<0.\end{cases}
\end{equation}
These are suitable extensions {of the considered} distributions to superspace, as we will show throughout this paper. It is easy to check that a property similar to Proposition \ref{SplitAnalFunc} $i)$ holds for the above definitions. i.e.\
\begin{align}\label{DelSplitt}
\del^{(k)}(a+b) &= \sum_{j=0}^{n} \frac{{\bf b}^j}{j!}\, \del^{(k+j)}(a+b_0), & k&\in\N-2,
\end{align}
where $a=a_0+{\bf a}$, $b=b_0+{\bf b}$ are elements in $C^\infty(\R^m)\otimes \mathfrak{G}_{2n}^{(ev)}$.

Let us prove now some important properties of the $\del$-distribution in superspace. 

\begin{pro}\label{delSup}
Let  $g=g_0+{\bf g}\in C^\infty(\R^m)\otimes \mathfrak{G}_{2n}^{(ev)}$ be such that $\pa_{\underline x}[g_0]\neq0$ on $g_0^{-1}(0)$. Then, for $j\in \N\cup \{0\}$ it holds {that}
\begin{itemize}
\item[$i)$] $g^j \del^{(j)}(g) = (-1)^j j!\, \del(g)$, 
\item[$ii)$]  $g^{j+1} \del^{(j)}(g)=0$.
\end{itemize}
\end{pro}
\pf
Using Proposition \ref{1del} we get,  
\begin{align*}
g^j \del^{(j)}(g) &= (g_0+{\bf g})^j\, \del^{(j)}(g_0+{\bf g}) = \left[ \sum_{k=0}^{j} \binom{j}{k} g_0^{j-k} {\bf g}^k\right] \left[ \sum_{\ell=0}^{n} \frac{{\bf g}^{\ell}}{\ell !} \, \del^{(j+\ell)}(g_0)\right]\\
&= \sum_{p=0}^{n} \sum_{k=0}^{\min(j,p)}  \binom{j}{k} \frac{{\bf g}^p}{(p-k)!} \, g_0^{j-k}  \,\del^{(j+p-k)}(g_0)\\
&= \sum_{p=0}^{n} {\bf g}^p \sum_{k=0}^{\min(j,p)} (-1)^{j-k} \binom{j}{k} \frac{(j-k)!}{(p-k)!} \binom{j+p-k}{j-k}\del^{(p)}(g_0)\\
&= (-1)^j \sum_{p=0}^{n} \frac{{\bf g}^p}{p!} \del^{(p)}(g_0) \left[\sum_{k=0}^{\min(j,p)} (-1)^k \binom{j}{k}  \frac{(j+p-k)!}{(p-k)!}\right].
\end{align*}
Writing $(j+p-k)! = (-1)^k\frac{(j+p)!}{(-j-p)_k}$ and $(p-k)!=(-1)^k\frac{(p)!}{(-p)_k}$ we get
\[\sum_{k=0}^{\min(j,p)} (-1)^k \binom{j}{k}  \frac{(j+p-k)!}{(p-k)!} = \frac{(j+p)!}{p!} \sum_{k=0}^{\min(j,p)} (-1)^k \binom{j}{k} \frac{(-p)_k}{(-j-p)_k}=\frac{(j+p)!}{p!} {}_2F_1(-j, -p, -p-j, 1),\]
where  ${}_2F_1(a, b, c, z)$ denotes the hypergeometric function {with} parameters $a,b,c$ in the variable $z$, see \cite[p.~64]{MR1688958}. Using the Chu-Vandermonde identity (see \cite[p.~67]{MR1688958}) we obtain ${}_2F_1(-j, -p, -p-j, 1) =\frac{j!p!}{(j+p)!}$. 
Whence, $\sum_{k=0}^{\min(j,p)} (-1)^k \binom{j}{k}  \frac{(j+p-k)!}{(p-k)!} = j!$, and as a consequence,
\[g^j \del^{(j)}(g)= (-1)^j  j! \sum_{p=0}^{n} \frac{{\bf g}^p}{p!} \del^{(p)}(g_0)= (-1)^j j!\, \del(g).\]

\noindent For the {proof of} $ii)$ it follows from Proposition \ref{1del} that
\[g\del(g)=  \sum_{j=0}^{n} \frac{{\bf g}^j}{j!} g_0 \del^{(j)}(g_0) +  \sum_{j=0}^{n-1} \frac{{\bf g}^{j+1}}{j!} \del^{(j)}(g_0) =  -  \sum_{j=1}^{n} \frac{{\bf g}^j}{(j-1)!} \del^{(j-1)}(g_0) +  \sum_{j=0}^{n-1} \frac{{\bf g}^{j+1}}{j!} \del^{(j)}(g_0)=0.\]
Hence, using $i)$ we have {that} $g^{j+1} \del^{(j)}(g)=(-1)^j j!\, g \del(g)=0$. $\hfill\square$

\begin{pro}\label{delnul}
Let $g=g_0+{\bf g}$  and $h=h_0+{\bf h}$ be elements in $C^\infty(\R^m)\otimes \mathfrak{G}_{2n}^{(ev)}$  where $g_0$  and $h_0$ are their  real valued bodies and ${\bf g}$ resp.\ ${\bf h}$ their nilpotent parts. Let us assume that $\pa_{\underline x}[g_0]\neq0$ on $g_0^{-1}(0)$ and $h_0>0$ in $\R^m$. Then,
\[\del(hg)=\frac{\del(g)}{h}.\]
\end{pro}
\pf
We first prove the result for every real valued function $h_0>0$. From Proposition \ref{ProMultH} we obtain
\[\del(h_0 g) = \del(h_0 g_0 + h_0{\bf g})= \sum_{j=0}^{n} \frac{h_0^j {\bf g}^j}{j!} \, \del^{(j)}(h_0 g_0) =  \sum_{j=0}^{n} \frac{{\bf g}^j}{j!} \, \frac{\del^{(j)}( g_0)}{h_0} = \frac{\del( g)}{h_0}.\]
Writing $\frac{{\bf h}}{h_0}={\bf L}$, we get for $h=h_0+{\bf h}$ that $\del(hg)=\del\left(h_0(g+{\bf L}g)\right)=\frac{\del\left((g+{\bf L}g)\right)}{h_0}.$
{It thus} suffices to prove that for ${\bf L}$ nilpotent the equality $\displaystyle \del\left((g+{\bf L}g)\right)=\frac{\del(g)}{1+{\bf L}}$ holds. Finally using (\ref{DelSplitt}) and Proposition \ref{delSup} $i)$ we get 
\[\del(g+{\bf L}g) = \sum_{j=0}^{n} \frac{{\bf L}^j g^j}{j!}\, \del^{(j)}(g)= \left(\sum_{j=0}^{n} (-1)^j j! \frac{{\bf L}^j}{j!}\right) \, \del(g) =\del(g) \sum_{j=0}^{n} (-1)^j {\bf L}^j=\frac{\del(g)}{1+{\bf L}}.
\]
$\hfill\square$
\begin{pro}\label{minus}
Let $g=g_0+{\bf g}\in C^\infty(\R^m)\otimes \mathfrak{G}_{2n}^{(ev)}$ be such  that $\pa_{\underline x}[g_0]\neq0$ on $g_0^{-1}(0)$. Hence $\del^{(j)}(-g)=(-1)^j\del^{(j)}(g)$ for every $j\in\N\cup\{0\}$.
\end{pro}
\pf This directly follows from the {corresponding} property in real analysis; $\del^{(j)}(-x)=(-1)^j\del^{(j)}(x)$. Indeed,
\[\del^{(j)}(-g)=\sum_{k=0}^n\frac{(-{\bf g})^k}{k!} \del^{(j+k)}(-g_0)=(-1)^j \sum_{k=0}^n\frac{({\bf g})^k}{k!} \del^{(j+k)}(g_0)=(-1)^j\del^{(j)}(g).\]
$\hfill\square$

\section{Integration in superspace}\label{Sec4}
In this section we define general domain and surface integration in superspace using the above definitions for the Heaviside and Dirac distributions. This {form} of integration turns out to be an easy and powerful formalism which has as natural antecedent in the real case.

Indeed, let $\Om\inc \R^m$ be a domain defined {by means of} a function $g_0\in C(\R^m)$ {as}  $\Om=\{\underline{x}\in\R^m: g_0(x)<0\}$. The characteristic function of $\Om$ is given by $H(-g_0)$; {this} easily leads to the following expression for the integration over $\Om$
\begin{equation}\label{RealDomInt}
\int_\Om f(\underline{x}) \,dV_{\underline{x}}= \int_{\R^m} H(-g_0(\underline{x})) f(\underline{x})\, dV_{\underline{x}}.
\end{equation}
The function $g_0$ is called the defining {\it phase function} of the domain $\Om$.

Let us now consider $g_0\in C^\infty(\R^m)$ such that $\pa_{\underline{x}}[g_0]\neq 0$  on the $(m-1)$-surface $\Gam:=g_0^{-1}(0)=\{\underline{w}\in\R^m: g_0(\underline{w})=0\}$. Then the non-oriented integral of a function $f$ over $\Gam$ can be written as the simple layer integral 
\begin{equation}\label{RealSurfInt01}
 \int_{\R^m} \del(g_0(\underline{x}))\, |\pa_{\underline{x}}[g_0](x)|\, f(\underline{x})\; dV_{\underline{x}}=\int_\R \del(t)\left(\int_{g_0^{-1}(t)} f(\underline{w})\, dS_{\underline{w}}\right)\, dt = \int_\Gam f(\underline{w})\, dS_{\underline{w}},
\end{equation}
where $dS_{\underline{w}}$ is the corresponding Lebesgue measure; see Theorem 6.1.5  \cite[p.~136]{MR1996773}. Observe also that the exterior normal vector to $\Gam$ in a point $\underline{w}\in \Gam$ is given by $\displaystyle n(\underline{w})=\frac{\pa_{\underline{x}}[g_0](\underline{w})}{|\pa_{\underline{x}}[g_0](\underline{w})|}$.  This leads to the following formula for the oriented surface integral
\begin{equation}\label{RealSurfInt1}
 \int_\Gam n(\underline{w}) f(\underline{w})\, dS_{\underline{w}} =  \int_{\R^m} \del(g_0(\underline{x}))\, \pa_{\underline{x}}[g_0](x)\, f(\underline{x})\;dV_{\underline{x}}.
\end{equation}

Formulas (\ref{RealDomInt}), (\ref{RealSurfInt01}) and (\ref{RealSurfInt1}) share a very important characteristic: they describe integrals over specific domains and surfaces as integrals over the whole space $\R^m$ depending only on the defining function $g_0$. In other words, they show the transition of the concept of integral as a functional depending on a set of points of $\R^m$ to a functional depending on a fixed phase function $g_0$. 

This last approach will be used to define domain and surface integration in superspace using formulas {which are similar} to (\ref{RealDomInt}),(\ref{RealSurfInt01}),(\ref{RealSurfInt1}). 
Given an analytic even superfunction $g\in C^\infty(\R^m)\otimes \mathfrak{G}_{2n}^{(ev)}$, one may consider the superdistributions $H(-g)$ and $\del(g)$ as the formal "characteristic functions" for the domain and surface associated to $g$ respectively. As in the classical case, the superfunction $g$ defining a domain in superspace is called a phase function. 

\begin{example}\label{ExSS}
The supersphere $R\Sa^{2m-1|2n}$ and {the corresponding} superball $R\B^{2m|2n}$ of radius $R>0$ are associated to the superfunction
\[-g({\bf x})= {\bf x}^2+R^2=R^2-|\underline{x}|^2+\underline{x}\p^2=R^2 -\sum_{j=1}^m x_j^2+\sum_{j=1}^{2n} x\p_{2j-1} x\p_{2j}.\]
The Dirac distribution corresponding to the superspshere $R\Sa^{2m-1|2n}$ is
\[\del({\bf x}^2+R^2)=\sum_{j=0}^n \frac{\underline{x}\p^{\,2j}}{j!} \;\del^{(j)}(R^2-|\underline{x}|^2).\]
The Heaviside distribution corresponding to the superball $R\B^{2m|2n}$ is
\[H({\bf x}^2+R^2)=H(R^2-|\underline{x}|^2)+\sum_{j=1}^{n} \frac{\underline{x}\p^{\,2j}}{j!} \,\del^{(j-1)}(R^2-|\underline{x}|^2).\]
\end{example}

\subsection{Domain integrals in superspace}
As mentioned before, our approach considers domains in superspace $\Om_{m|2n}$ given {by} characteristic functions of the form $H(-g)$ where $g({\bf x})=g_0(\underline{x})+{\bf g}(\underline{x}, \underline{x}\p)\in C^\infty(\R^m)\otimes \mathfrak{G}_{2n}^{(ev)}$. In this sense, $\Om_{m|2n}$ plays the same r\^ole in superanalysis as its associated real domain $\Om_{m|0}:=\{\underline{x}\in\R^m: g_0(\underline{x})<0\}$ in classical analysis.
\begin{defi}\label{DomIntSupS}
Let $\Om_{m|2n}$ be a domain in superspace (defined as before) satisfying the following two conditions:
\begin{itemize}
\item the associated real domain $\Om_{m|0}$ has compact closure;
\item the body $g_0$ of the defining superfunction $g$ is such that $\pa_{\underline x}[g_0]\neq0$ on $g_0^{-1}(0)$.
\end{itemize}
The integral over $\Om_{m|2n}$ {then} is defined as the functional $\int_{\Om_{m|2n}}:C^{n-1}\left(\ba{\Om_{m|0}}\right)\otimes \mathfrak{G}_{2n}\fd \R$ given by
\begin{equation}\label{DomSupInt}
\int_{\Om_{m|2n}} F= \int_{\R^{m|2n}} H(-g) F, \hspace{2cm} F\in C^{n-1}\left(\ba{\Om_{m|0}}\right)\otimes \mathfrak{G}_{2n}.
\end{equation}
\end{defi}

The evaluation of the expression (\ref{DomSupInt}) requires the integration of smooth functions on the real domain $\Om_{m|0}$ {to be possible}. This is guaranteed by the first condition imposed on the super-domain $\Om_{m|2n}$. On the other hand, if ${\bf g}$ is not identically zero, the above definition {also} involves the action of the Dirac distribution {on} $g_0$, see (\ref{HaevSupS}). For that reason, we restrict our analysis to the case {where} $\pa_{\underline x}[g_0]\neq0$ on $g_0^{-1}(0)$, {in order to ensure that this action} is well defined.

The most simple examples {for illustrating} the use of Definition \ref{DomIntSupS} {correspond to the cases} $g=g_0\in C^\infty(\R^m)$ or $g=-{\bf x}^2-R^2$; i.e. integration over real domains or over the superball respectively. The integral (\ref{DomSupInt}) {then} is given by
\[\int_\Om \int_B F dV_{\underline x},\hspace{.5cm} \mbox{ and } \hspace{.5cm} \int_{\R^{m|2n}} H({\bf x}^2+R^2) F,\]
respectively. These are the two particular cases that have been treated in the literature, see \cite{MR2539324, MR2521367}. The superball (and the supersphere) will {also} be used in this manuscript {as an illustrative example}. In the next sections we work with integrals over other super-domains such a as a super-paraboloid and a super-hyperboloid.
\begin{pro}\label{VolSS}
The volume of the superball $R\,\B^{m|2n}$ of radius $R>0$ is {given  by}
\[vol(R\,\B^{m|2n})=\begin{cases}\displaystyle\frac{\pi^{M/2}}{\Gam(\frac{M}{2}+1)}R^M, & M\notin -2\N,\\ 0, & M\in -2\N,\end{cases}\]
where $M=m-2n$ ($m\neq0$) is the corresponding superdimension.
\end{pro}
\pf
The volume of the $R\,\B^{m|2n}$ is obtained by integrating the {function} $F\equiv 1$ over $R\,\B^{m|2n}$ following Definition \ref{DomIntSupS}; i.e. 
\[vol(R\,\B^{m|2n})=\int_{R\,\B^{m|2n}} 1= \int_{\R^m} \int_B H(R^2-|\underline{x}|^2+\underline{x}\p^2)\, dV_{\underline{x}},\]
where $H(R^2-|\underline{x}|^2+\underline{x}\p^2)=H(R^2-|\underline{x}|^2)+\sum_{j=1}^{n} \frac{\underline{x}\p^{\,2j}}{j!} \,\del^{(j-1)}(R^2-|\underline{x}|^2)$.
Since $\underline{x}\p^{\,2n} = n! {x}\p_1\cdots x\p_{2n}$, we obtain $\displaystyle \int_B H(R^2-|\underline{x}|^2+\underline{x}\p^2)= \pi^{-n}\,\del^{(n-1)}(R^2-|\underline{x}|^2)$ and in {consequently},
\begin{equation}\label{VolSB}
vol(R\,\B^{m|2n})= \pi^{-n}\,\int_{\R^m} \del^{(n-1)}(R^2-|\underline{x}|^2) \, dV_{\underline{x}}.
\end{equation}
{Effectuating} the change of variables $\underline{x}= r \underline{w}$, where $0<r< \infty$ and $\underline{w}\in\Sa^{m-1}:=\{\underline{x}\in \R^m: \underline{x}^2=-1\}$, we get $dV_{\underline{x}}=r^{m-1} dr\, dS_{\underline{w}}$. Hence, 
\[vol(R\,\B^{m|2n})=\pi^{-n} \int_{\Sa^{m-1}} \left(\int_0^\infty \del^{(n-1)}(R^2-r^2)\, r^{m-1} \,dr \right) \, dS_{\underline{w}}= \pi^{-n}\,A_m \int_0^\infty \del^{(n-1)}(R^2-r^2)\, r^{m-1} \,dr, \]
where $A_m= \int_{\Sa^{m-1}} dS_{\underline{w}}=\frac{2\pi^{m/2}}{\Gam(\frac{m}{2})}$ is the area of the unit sphere $\Sa^{m-1}$ in $\R^m$. Changing variables again, now through $t=R^2-r^2$, we obtain $r=\left(R^2-t\right)^{1/2}$ and $dr=-\frac{1}{2} \left(R^2-t\right)^{-1/2}dt$. Then,  it holds that
\begin{align}
vol(R\,\B^{m|2n}) &= \frac{\pi^{-n}\,A_m}{2} \int_{-\infty}^{R^2} \del^{(n-1)}(t) \left(R^2-t\right)^{\frac{m}{2}-1}\; dt \nonumber \\
&= (-1)^{n-1} \frac{\pi^{-n}\,A_m}{2} \; \frac{d^{n-1}}{dt^{n-1}} \left[\left(R^2-t\right)^{\frac{m}{2}-1} \right] \Bigg|_{t=0}\nonumber \\
&= \frac{\pi^{-n}\,A_m}{2} \prod_{j=1}^{n-1}\left(\frac{m}{2}-j\right)R^{{m}-2n}, \label{prod}
\end{align}
where for the special cases $n=0,1$ we are considering $\prod_{j=1}^{-1}\left(\frac{m}{2}-j\right):=\frac{2}{m}$ and $\prod_{j=1}^{0}\left(\frac{m}{2}-j\right):=1$.
Since we consider $m\neq 0$, it is easily seen that 
\[\prod_{j=1}^{n-1}\left(\frac{m}{2}-j\right)=\begin{cases}\frac{\Gam(\frac{m}{2})}{\Gam(\frac{m}{2}-n+1)} & \frac{M}{2}\notin -\N,\\ 0& \frac{M}{2}\in -\N\end{cases}\]
{whence substitution} in (\ref{prod}) completes the proof. $\hfill\square$

\begin{remark}
Proposition \ref{VolSS} provides a suitable extension to superspace for the volume of the ball $R\B^{m}:=\{\underline{x}\in \R^m: |\underline{x}|<R\}$  ($R>0$). We recall that in the case $n=0$ the volume of $R\B^{m}$ equals $\frac{\pi^{m/2}}{\Gam(\frac{m}{2}+1)}R^m$.
\end{remark}

In real analysis {the} choice {of} the function $g_0$ {defining} a certain domain $\Om_{m|0}\inc\R^m$ {is not unique}. Indeed, for every function $h_0\in C(\R^m)$ {with} $h_0>0$, the function $h_0g_0$ defines the same domain as $g_0$, i.e.
\[\Om_{m|0}=\{\underline{x}\in \R^m: g_0(\underline{x})<0\}=\{\underline{x}\in \R^m: h_0(\underline{x})g_0(\underline{x})<0\}.\]
A simple example is the unitary ball $\B\inc \R^m$ which can be described by 
\[|\underline{x}|-1<0, \hspace{.3cm} \mbox{ or } \hspace{.3cm}  -\underline{x}^2-1<0, \hspace{1cm}  \mbox{ {since}, } \hspace{.2cm}  -\underline{x}^2-1= (|\underline{x}|+1)(|\underline{x}|-1) .\]
However, integration over $\Om_{m|0}$ remains  independent of the choice of the function $g_0$ that defines $\Om_{m|0}$. Indeed, in real analysis it is easily seen that $H(-g_0)=H(-h_0g_0)$ for $h_0>0$. This property remains valid in superspace. 
\begin{pro}\label{DomInv}
Let $g=g_0+{\bf g}$  and $h=h_0+{\bf h}$ be elements in $C^\infty(\R^m)\otimes \mathfrak{G}_{2n}^{(ev)}$  where $g_0$  and $h_0$ are their  real valued bodies and ${\bf g}$ resp.\ ${\bf h}$ their nilpotent parts. Let us assume that $\pa_{\underline x}[g_0]\neq0$ on $g_0^{-1}(0)$ and $h_0>0$ in $\R^m$. Then $H(hg)=H(g)$.
\end{pro}
\pf
We first prove that $H(h_0g)=H(g)$. By Proposition \ref{ProMultH} we obtain $(h_0{\bf g})^j\del^{(j-1)}(h_0g_0)={\bf g}^j\del^{(j-1)}(g_0)$ for $j\in\N$. Hence,
\[H(h_0g)=H(h_0g_0)+\sum_{j=1}^{n} \frac{(h_0{\bf g})^j}{j!} \del^{(j-1)}(h_0g_0) = H(g_0)+\sum_{j=1}^{n} \frac{{\bf g}^j}{j!} \del^{(j-1)}(g_0)=H(g).\]
If we write ${\bf L}=\frac{{\bf h}}{h_0}$, we get for $h=h_0+{\bf h}$ that $H(hg)=H(h_0g+{\bf h}g)=H(g+{\bf L}g)$, {whence} it suffices to prove that $H(g+{\bf L}g)=H(g)$. Using (\ref{DelSplitt}) and Proposition \ref{delSup} $ii)$ we obtain
\[H(g+{\bf L}g)=H(g)+\sum_{j=1}^{n} \frac{{\bf L}^j }{j!}\, g^j \del^{(j-1)}(g)= H(g). \]
$\hfill\square$

This allows {us} to define the notion of a pair of superfunctions that define the same domain (or surface).
\begin{defi}
Let $g,\fhi\in C^\infty(\R^m)\otimes \mathfrak{G}_{2n}^{(ev)}$ be two superfunctions such that $\pa_{\underline x}[g_0]\neq0$ on $g_0^{-1}(0)$ and $\pa_{\underline x}[\fhi_0]\neq0$ on $\fhi_0^{-1}(0)$. They are said to define the same domain (or surface) in superspace if there exists a function $h\in C^\infty(\R^m)\otimes \mathfrak{G}_{2n}^{(ev)}$ with $h_0>0$ in $\R^m$ such that $\fhi=hg$.
\end{defi}

\begin{remark}
Proposition \ref{DomInv} shows that the integral over $\Om_{m|2n}$ defined in (\ref{DomSupInt}) does not depend on the choice of the superfunction $g$ defining $\Om_{m|2n}$. The example of the superball $R\,\B^{m|2n}$ of radius $R>0$ illustrates very well this property. Indeed, the domain $R\,\B^{m|2n}$ can be defined by means of any of the two superfunctions $|{\bf x}|-R$ or $-{\bf x}^2-R^2$.
Both definitions for $R\,\B^{m|2n}$ can be used in (\ref{DomSupInt}) without changing the result of the integration since $-{\bf x}^2-R^2=(|{\bf x}|+R)(|{\bf x}|-R)$ where
\[h({\bf x})=|{\bf x}|+R= R+\sum_{j=0}^n \frac{(-1)^j \underline{x\p}^{\,2j}}{j!} \, \frac{\Gam(\frac{3}{2})}{\Gam(\frac{3}{2}-j)} |\underline{x}|^{1-2j}, \hspace{.3cm} \mbox{ and } \hspace{.3cm} h_0(\underline{x})=R+|\underline{x}|>0.\]
\end{remark}

\subsection{Surface integrals in superspace}
{Similarly to} the case of super domains, we define a surface $\Gam_{m-1|2n}$ in superspace by means of $\del(g)$ where $g({\bf x})=g_0(\underline{x})+{\bf g}(\underline{x}, \underline{x}\p)\in C^\infty(\R^m)\otimes \mathfrak{G}_{2n}^{(ev)}$. If $\Om_{m|2n}$ is the super domain associated to $g$ as in Definition \ref{DomIntSupS}, then we say that $\Gam_{m-1|2n}$ is the boundary of $\Om_{m|2n}$ and denote it by $\Gam_{m-1|2n}:=\pa\Om_{m|2n}$. This way, $\Gam_{m-1|2n}$ plays the same r\^ole in superspace as its real surface $\Gam_{m-1|0}:=\pa\Om_{m|0}=\{\underline{x}\in \R^m: g_0(\underline{x})=0\}$ in classical analysis. 

Based on the formulae (\ref{RealSurfInt01}) and (\ref{RealSurfInt1}) concerning the real case we now define the non-oriented and oriented surface integrals in superspace.
\begin{defi}\label{SurfIntSupS}
Let  $\Gam_{m-1|2n}$ be a surface in superspace (defined as before) satisfying the following two conditions:
\begin{itemize}
\item the associated real surface $\Gam_{m-1|0}\inc \R^m$ is a comapact set;
\item the body $g_0$ of the defining superfunction $g$ is such that $\pa_{\underline x}[g_0]\neq0$ on $g_0^{-1}(0)$.
\end{itemize}
The non-oriented and oriented surface integrals over $\Gam_{m-1|2n}$ {then} are defined as the following functionals  on $C^{n}\left(\Gam_{m-1|0}\right)\otimes \mathfrak{G}_{2n}$\begin{equation}\label{SurfIntNO}
\int_{\Gam_{m-1|2n}} F = \int_{\R^{m|2n}} \del(g) \, \big|\pa_{{\bf x}}[g]\big| \, F,\hspace{.8cm} \int_{\Gam_{m-1|2n}} \sigma_{\bf x}\, F = - \int_{\R^{m|2n}} \del(g) \, \pa_{{\bf x}}[g] \, F,
\end{equation}
respectively.
\end{defi}
\begin{remark}
Proposition \ref{minus} assures that the sign of the superfunction $g$ does not {play a r\^ole} in the non-oriented case.
\end{remark}


When $g=g_0$, the integrals (\ref{SurfIntNO})  reduce to the  product of the classical real surface integration  and the Berezin integral, i.e.,
\begin{equation}\label{RealTimBer}
\int_{\Gam_{m-1|0}} \int_B F(\underline{w}, \underline{x}\p)\, dS_{\underline{w}}, \hspace{.8cm} \int_{\Gam_{m-1|0}} \int_B n(\underline{w})F(\underline{w}, \underline{x}\p)\, dS_{\underline{w}},
\end{equation}
see formulae (\ref{RealSurfInt01})-(\ref{RealSurfInt1}). {We will now verify} that, when restricted to the supersphere $R\Sa^{2m-1|2n}$, the definition {of the} non-oriented surface integral still coincides with the one given in \cite{MR2539324}.

\begin{pro}
The non-oriented integral over the supersphere $R\Sa^{2m-1|2n}$ ($R>0$) can be written as
 \begin{equation}\label{IntSS}
 \int_{R\Sa^{m-1,2n}} F= 2\int_{\R^m} \int_B \del(R^2+{\bf x}^2) \,|{\bf x}| \,F({\bf x})\; dV_{\underline x}=2 R \int_{\R^{m}} \int_B \del({\bf x}^2+R^2)\, F({\bf x}) \; dV_{\underline x}.
 \end{equation}
\end{pro}
\pf
We first observe that $\pa_{\bf x}[R^2+{\bf x}^2]=2{\bf x}$.  Following (\ref{SurfIntNO}), the non-oriented surface integral of $F$ over the supersphere $R\Sa^{m-1,2n}$ is given by
\[\int_{R\Sa^{m-1,2n}} F= 2\int_{\R^m} \int_B \del(R^2+{\bf x}^2) \,|{\bf x}| \,F({\bf x})\; dV_{\underline x}.\]
Let us prove now that, as in the classical case, one can substitute $|{\bf x}|=R$ in the above expression. Consider the real distribution $G(t)=\del(R^2-t)t^{\frac{1}{2}}$, $t\geq 0$. Then, we can write in superspace
\[\del(R^2+{\bf x}^2) \,|{\bf x}|= \del(R^2+{\bf x}^2) \,(-{\bf x}^2)^{\frac{1}{2}}=G(-{\bf x}^2)=G(|\underline{x}|^2-\underline{x}\p^{\,2})=\sum_{j=0}^n \frac{\left(-\underline{x}\p^{\,2}\right)^j}{j!} \,G^{(j)}(|\underline{x}|^2).\]
{However}, in the distributional sense, it is easily seen that  $G(t)=\del(R^2-t)t^{\frac{1}{2}}=R\, \del(t-R^2)$, {whence also} $G^{(j)}(t)=R\del^{(j)}(t-R^2)$, $j=0, \ldots, n$. Substituting this in the above formula we obtain,
\[\del(R^2+{\bf x}^2) \,|{\bf x}|=R\sum_{j=0}^n \frac{\left(-\underline{x}\p^{\,2}\right)^j}{j!} \del^{(j)}(|\underline{x}|^2-R^2)=R\, \del(R^2+{\bf x}^2), \]
which completes the proof.  $\hfill\square$
The non-oriented integration over the supersphere (\ref{IntSS}), and the integration of superfunctions over real surfaces (\ref{RealTimBer}), have been studied in the literature,  see e.g.\ \cite{MR2539324, MR2521367}. In particular, (\ref{IntSS}) was proven to be an extension of the Pizzetti formula for polynomials. The simplest example for application of the Pizzetti formulas is obtained when {integrating} the function $F\equiv 1$; which leads to the surface area of the supersphere $R\Sa^{m-1,2n}$. Such an area will be computed next in order to show the simplest example for this kind of integration. As mentioned before, new examples regarding to a super-paraboloid and a super-hyperboloid will be treated in the next section.
\begin{pro}
The surface area of the supersphere $R\Sa^{m-1,2n}$ of radius $R>0$ is {given by}
\[area(R\,\Sa^{m-1|2n})=\begin{cases}\displaystyle\frac{2\pi^{M/2}}{\Gam(\frac{M}{2})}R^{M-1}, & M\notin -2\N+2,\\ 0 & M\in -2\N+2, \end{cases}\]
where $M=m-2n$ ($m\neq0$) is the corresponding superdimension.
\end{pro}
\pf
Using (\ref{IntSS}), the area of the surface $R\Sa^{m-1,2n}$ is given by
 \[area(R\Sa^{m-1,2n}) =  \int_{R\Sa^{m-1,2n}} 1= 2 R \int_{\R^{m}} \int_B \del({\bf x}^2+R^2) \; dV_{\underline x},\]
 where $\del({\bf x}^2+R^2)=\sum_{j=0}^n \frac{\underline{x}\p^{\,2j}}{j!} \del^{(j)}(R^2-|\underline{x}|^2)$ implies $\int_B\del({\bf x}^2+R^2) =\pi^{-n} \,\del^{(n)}(R^2-|\underline{x}|^2)$. By the {co-ordinate} changes $\underline{x}=r\underline{w}$, $\underline{w}\in \Sa^{m-1}$ and $r=t^{\frac{1}{2}}$  we get 
  \begin{align}
 area(R\Sa^{m-1,2n}) &=  2 R\pi^{-n} \int_{\R^{m}} \del^{(n)}(R^2-|\underline{x}|^2) \; dV_{\underline x}\nonumber\\
 &= 2R\pi^{-n}  \int_{\Sa^{m-1}} \left(\int_0^\infty  \del^{(n)}(R^2-r^2) \, r^{m-1} \; dr \right) \; dS_{ \underline{w}}\nonumber\\
 &= \pi^{-n} \,A_m \, R \int_0^\infty  \del^{(n)}(R^2-t) t^{\frac{m}{2}-1} \; dt \nonumber\\
 &= \pi^{-n} \,A_m \, R \;\frac{d^n}{dt^n} \left[t^{\frac{m}{2}-1}\right] \bigg|_{t=R^2}\nonumber\\
 &= \pi^{-n} \,A_m \, \prod_{j=1}^n\left(\frac{m}{2}-j\right) R^{{m}-2n-1},\label{ArMult}
 \end{align}
where for the special case $n=0$ we {put by convention} $\prod_{j=1}^0\left(\frac{m}{2}-j\right):=1$. Since we consider $m\neq 0$, it is easily seen that 
\[\prod_{j=1}^{n}\left(\frac{m}{2}-j\right)=\begin{cases}\frac{\Gam(\frac{m}{2})}{\Gam(\frac{m}{2}-n)} & \frac{M}{2}\notin -\N+1,\\ 0& \frac{M}{2}\in -\N+1\end{cases}\]
{whence substitution}  in (\ref{prod})  completes the proof. $\hfill\square$

 As for domain integrals, we can prove that Definition \ref{SurfIntSupS} does not depend on the choice of the defining superfunction $g$ for the surface $\Gam_{m-1|2n}$.
 
\begin{pro}
Let $g=g_0+{\bf g}$  and $h=h_0+{\bf h}$ be elements in $C^\infty(\R^m)\otimes \mathfrak{G}_{2n}^{(ev)}$  {having} $g_0$ and $h_0$ {as} their  real valued bodies and ${\bf g}$ resp.\ ${\bf h}$ their nilpotent parts. Let us assume that $\pa_{\underline x}[g_0]\neq0$ on $g_0^{-1}(0)$ and $h_0>0$ in $\R^m$. Then,
\[i)\;\; \del(hg)\pa_{\bf x}[hg]=\del(g)\pa_{\bf x}[g], \hspace{2cm} ii)\;\; \del(hg) |\pa_{\bf x}[hg]|=\del(g)|\pa_{\bf x}[g]|. \]
\end{pro}
\pf Using  Propositions \ref{delSup} and \ref{delnul} we get $\del(hg)\pa_{\bf x}[hg]= \frac{\del(g)}{h} \left(\pa_{\bf x}[h]g+h\pa_{\bf x}[g]\right)=\del(g)\pa_{\bf x}[g]$, which proves $i)$. 
In addition, by Propositions \ref{PropMod} and \ref{delnul} we get
\[ \del(hg) |\pa_{\bf x}[hg]| =  \del(hg) \big|\pa_{\bf x}[h]g+h\pa_{\bf x}[g]\big|=\frac{\del(g)}{h} \bigg[h|\pa_{\bf x}[g]|+gF\big(h\pa_{\bf x}[g], \pa_{\bf x}[h], g\big)\bigg]=\del(g)|\pa_{\bf x}[g]|,\]
which proves $ii)$.
$\hfill\square$

\section{Other examples and applications}
In this section we study {some more} examples of integration in superspace over domains and surfaces. To that end, let us observe first that, as in the classical case, it is possible to define intersections amongst domains {as well as}  between a surface and domains in superspace.

Indeed, let $\Om_j$ ($j=1,\ldots, k$) be domains in superspace defined by the characteristic functions $H(-g_j)$ with $g_j({\bf x})=g_{j,0}(\underline{x})+{\bf g}_j(\underline{x}, \underline{x}\p)\in C^\infty(\R^m)\otimes \mathfrak{G}_{2n}^{(ev)}$. The intersection $\Om_{m|2n}=\cap_{j=1}^k \Om_k$ is naturally defined as the domain with characteristic function $H(-g_1)\ldots H(-g_k)$. If we also consider a surface $\Gam_{m-1|2n}$ defined by $\del(g)$ with $g({\bf x})=g_{0}(\underline{x})+{\bf g}(\underline{x}, \underline{x}\p)\in C^\infty(\R^m)\otimes \mathfrak{G}_{2n}^{(ev)}$, the intersection $\Gam_{m-1|2n}\cap \Om_{m|2n}$ is defined by the characteristic function $\del(g)H(-g_1)\ldots H(-g_k)$. The domain $\Om_{m|2n}=\cap_{j=1}^k \Om_k$ plays the same r\^ole in superspace as its associated region $\Om_{m|0}:=\bigcap_{j=1}^k \left\{\underline{x}\in\R^m: g_{j,0}(\underline{x})<0\right\}$ in $\R^m$. In the same way, $\Gam_{m-1|2n}\cap \Om_{m|2n}$ is the super-analogue of the intersection of the ($m-1$)-surface $\Gam_{m-1|0}=g_0^{-1}(0)$ with the region $\Om_{m|0}$.

{Now} assume that $\Om_{m|0}$ has a compact closure and the body functions $g_0, g_{1,0}, \ldots, g_{k,0}$ have non vanishing gradients on the respective surfaces $g_0^{-1}(0), g_{1,0}^{-1}(0), \ldots, g_{k,0}^{-1}(0)$. Hence, Definitions \ref{DomIntSupS} and \ref{SurfIntSupS} extend to $\Om_{m|2n}=\cap_{j=1}^k \Om_k$ and $\Gam_{m-1|2n}\cap \Om_{m|2n}$ respectively by {putting}
\begin{align*}
\int_{\Om_{m|2n}} F&=\int_{\R^{m|2n}} H(-g_1)\ldots H(-g_k) F, \\
\int_{\Gam_{m-1|2n}\cap \Om_{m|2n}} F&= \int_{\R^{m|2n}} \del(g) \, \big|\pa_{{\bf x}}[g]\big| \, H(-g_1)\ldots H(-g_k) F,\\
 \int_{\Gam_{m-1|2n}\cap \Om_{m|2n}} \sigma_{\bf x}\, F &= - \int_{\R^{m|2n}} \del(g) \, \pa_{{\bf x}}[g] \, H(-g_1)\ldots H(-g_k) F.
\end{align*}

\subsection{Super-paraboloid of revolution}
Let us consider the paraboloid of revolution in 3 real dimensions defined by 
\[x_3=x_1^2+x_2^2=-\underline{\hat{x}}^2 \hspace{.5cm}\mbox{ where } \hspace{.5cm} \underline{\hat{x}}=x_1e_1+x_2e_2.\]
We define its extension to superspace by means of the superfunction
\[g({\bf x})=-{\hat{{\bf x}}}^2-x_m=\sum_{j=1}^{m-1} x_j^2 - \sum_{j=1}^n x\p_{2j-1}x\p_{2j}-x_m=|\underline{\hat{x}}|^2-x_m-\underline{{x}}\p^{\,2} \hspace{.5cm} (m\geq 2),\]
where, in this case, $\underline{\hat{x}}=\underline{{x}}-x_me_m$ and $\hat{{\bf x}}=\underline{\hat{x}}+ \underline{{x}}\p={\bf x}- x_me_m$.

The set {$\{\underline{x}\in\R^m: |\underline{\hat x}|^2-x_m \leq 0\}$ clearly is non} compact. {However, its} intersection with the region $\{\underline{x}\in\R^m: x_m\in[0,h] \}$ ($h>0$) gives a compact subset composed by the interior and boundary of the paraboloid $x_m=|\underline{\hat x}|^2$ with height $h>0$. This means that we can integrate over the domain (and surface) defined by the superfunction $g$ in superspace with the restriction $x_m<h$. This object will be called {the} {\it super-paraboloid of revolution of height $h>0$} and {is} denoted by $SP^{m|2n}_h$. More precisely, the domain and surface associated to $SP^{m|2n}_h$ are given by the characteristic functions
\[H(-g) H(h-x_m), \hspace{.5cm} \mbox{ {and} } \hspace{.5cm}\del(g)H(h-x_m),\]
respectively.
\begin{pro}
The volume of  $SP^{m|2n}_h$ is {given by}
\begin{equation}\label{VolSP}
vol(SP^{m|2n}_h)= \begin{cases}\displaystyle\frac{\pi^{\frac{M-1}{2}}}{\Gam(\frac{M+3}{2})}\, h^{\frac{M+1}{2}}, & M\notin -2\N+1,\\ 0 & M\in -2\N+1. \end{cases}
\end{equation}
\end{pro}
\pf
Observe that
\[vol(SP^{m|2n}_h)=\int_{\R^{m|2n}} H(-g)H(h-x_m)=\int_{0}^h \left(\int_{\R^{m-1}}\int_B H(-g)\; dV_{\underline{\hat{x}}}\right)dx_m.\]
Because $H(-g)=H(x_m-|\underline{\hat{x}}|^2+\underline{{x}}\p^{\,2})=\sum_{j=0}^n \frac{\underline{{x}}\p^{\,2j}}{j!}\del^{(j-1)}(x_m-|\underline{\hat{x}}|^2)$ we have $\int_B H(-g)=\pi^{-n}\, \del^{(n-1)}(x_m-|\underline{\hat{x}}|^2)$. 
Hence, by formula (\ref{VolSB}) we get
\[\int_{\R^{m-1}}\int_B H(-g)\; dV_{\underline{\hat{x}}}=\pi^{-n}\, \int_{\R^{m-1}}  \del^{(n-1)}(x_m-|\underline{\hat{x}}|^2)\; dV_{\underline{\hat{x}}} = vol\left(x_m^{\frac{1}{2}} \, \B^{m-1|2n}\right),\]
which leads to
\[\int_{\R^{m-1}}\int_B H(-g)\; dV_{\underline{\hat{x}}}=\begin{cases}\displaystyle\frac{\pi^{\frac{M-1}{2}}}{\Gam(\frac{M+1}{2})}\,x_m^{\frac{M-1}{2}}, & M\notin -2\N+1,\\ 0 & M\in -2\N+1, \end{cases}\]
{whence} for $M\notin -2\N+1$ we obtain
\[vol(SP^{m|2n}_h) = \frac{\pi^{\frac{M-1}{2}}}{\Gam(\frac{M+1}{2})} \int_{0}^h x_m^{\frac{M-1}{2}} \, dx_m= \frac{\pi^{\frac{M-1}{2}}}{\Gam(\frac{M+1}{2})} \frac{h^{\frac{M+1}{2}}}{\frac{M+1}{2}}=\frac{\pi^{\frac{M-1}{2}}}{\Gam(\frac{M+3}{2})}\, h^{\frac{M+1}{2}}; \]
and $vol(SP^{m|2n}_h) =0$ for $M\in -2\N+1$. $\hfill\square$

\begin{remark}\label{CCVSP}
The above result is an extension of the volume formulae for the corresponding paraboloids in the known classical cases $m=2$, $n=0$ and $m=3$, $n=0$. 
\begin{itemize}
\item In the case $m=2$, $n=0$, the parabola $SP_h^{2|0}=\{(x_1,x_2)\in \R^2: x_1^2\leq x_2\leq h\}$ is known to have the volume (i.e.\ area in $\R^2$)
\[2h^{3/2}-\int_{-h^{\frac{1}{2}}}^{h^{\frac{1}{2}}} x_1^2\, dx_1=\frac{4}{3}h^{\frac{3}{2}}.\]
This is exactly the result obtained when substituting $M=2$ in (\ref{VolSP}).
\item In the case $m=3$, $n=0$, $SP_h^{3|0}=\{(x_1,x_2, x_3)\in \R^3: x_1^2+x_2^2\leq x_3\leq h\}$ is the body generated by the rotation of the curve $x_3=x_1^2$ around the $x_3$-axis from $x_1=0$ to $x_1=h^{\frac{1}{2}}$. The volume of $SP_h^{3|0}$ is known to be
\[2\pi \int_0^{h^\frac{1}{2}}x_1(h-x_1^2)\, dx_1=\frac{\pi}{2}h^2,\]
{coinciding with} the result obtained when {evaluating} (\ref{VolSP}) for $M=3$. \end{itemize}

\end{remark}
\begin{pro}
The surface area of $SP^{m|2n}_h$ for $M>1$ is {given by}
\begin{equation}\label{AreaSP}
area (SP^{m|2n}_h)= \frac{\pi^{\frac{M-1}{2}}}{\Gam\left(\frac{M+1}{2}\right)} h^{\frac{M-1}{2}} \; {}_2F_1\left(-\frac{1}{2},\frac{M-1}{2};\frac{M+1}{2};-4h\right),
\end{equation}
where ${}_2F_1(a, b;c;z)$ denotes the hypergeometric function, see \cite[p.~64]{MR1688958}.
\end{pro}
\pf
In this case,
\[area (SP^{m|2n}_h)=\int_{\R^{m|2n}} \del(g) \, \big|\pa_{{\bf x}}[g]\big|\, H(h-x_m)=\int_{0}^h \left(\int_{\R^{m-1}} \int_B \del(g) \, \big|\pa_{{\bf x}}[g]\big|\; dV_{\underline{\hat{x}}}\right)\, dx_m.\]
We recall that $\pa_{\bf x}=\pa_{{\hat{{\bf x}}}} -\pa_{x_m}e_m$. Then $\pa_{\bf x}[g]=(\pa_{{\hat{{\bf x}}}} -\pa_{x_m}e_m)(-{\hat{{\bf x}}}^2-x_m)=-\pa_{{\hat{{\bf x}}}}[{\hat{{\bf x}}}^2]+\pa_{x_m}[x_m]e_m=-2\hat{{\bf x}}+e_m$, {which} leads to $\big|\pa_{{\bf x}}[g]\big|=\left(-4{\hat{{\bf x}}}^2+1\right)^{\frac{1}{2}}$. 

\noindent Let us {now} consider the distribution $D(t)=\del(t-x_m)(4t+1)^{\frac{1}{2}}$, ($t>0$). The evaluation of $D$ in $-{\hat{{\bf x}}}^2=|\underline{\hat{x}}|^2-\underline{{x}}\p^{\,2}$ equals $ \del(g) \, \big|\pa_{{\bf x}}[g]\big|$ and is given by
\[D(-{\hat{{\bf x}}}^2)=\sum_{j=0}^n\frac{(-1)^j\underline{{x}}\p^{\,2j}}{j!} D^{(j)}\left(|\underline{\hat{x}}|^2\right), \hspace{.3cm}\mbox{ {implying} } \hspace{.3cm}\int_B D(-{\hat{{\bf x}}}^2)= (-1)^n\pi^{-n} D^{(n)}\left(|\underline{\hat{x}}|^2\right).\] 
Then,
\[area(SP^{m|2n}_h)=(-1)^n \pi^{-n}\, \int_0^h \left(\int_{\R^{m-1}} D^{(n)}\left(|\underline{\hat{x}}|^2\right) dV_{\underline{\hat{x}}} \right) \, dx_m.\]
{By} the change of variables $\underline{\hat{x}}=r \underline{w}$, $r>0$, $\underline{w}\in \Sa^{m-2}$ we get $dV_{\underline{\hat{x}}} = r^{m-2} dr\, dS_{{\underline{w}}}$ and
\begin{align*}
(-1)^n \pi^{-n} \int_{\R^{m-1}} D^{(n)}\left(|\underline{\hat{x}}|^2\right) dV_{\underline{\hat{x}}}  &=(-1)^n \pi^{-n} \int_{\Sa^{m-2}} \left(\int_{0}^{+\infty} D^{(n)}\left(r^2\right) r^{m-2}\, dr \right)\, dS_{{\underline{w}}}\\
&= \frac{(-1)^n \pi^{-n} A_{m-1}}{2}  \int_{0}^{+\infty}   D^{(n)}\left(t\right) t^{\frac{m-3}{2}}\, dt\\
&= \frac{\pi^{-n} A_{m-1}}{2}  \int_{0}^{+\infty}   D\left(t\right) \frac{d^n}{dt^n}\left[t^{\frac{m-3}{2}}\right]\, dt\\
&=\frac{\pi^{-n} A_{m-1}}{2}  \frac{\Gam\left(\frac{m-1}{2}\right)}{\Gam\left(\frac{m-1}{2}-n\right)}\int_{0}^{+\infty}   \del\left(t-x_m\right) (4t+1)^{\frac{1}{2}} t^{\frac{m-3}{2}-n}\, dt\\
&= \frac{\pi^{\frac{M-1}{2}}}{\Gam\left(\frac{M-1}{2}\right)} (4x_m+1)^{\frac{1}{2}} \, x_m^{\frac{M-3}{2}},
\end{align*}
{whence},
\[area(SP^{m|2n}_h)= \frac{\pi^{\frac{M-1}{2}}}{\Gam\left(\frac{M-1}{2}\right)}  \int_0^h (4x_m+1)^{\frac{1}{2}} \, x_m^{\frac{M-3}{2}} \, dx_m=\frac{\pi^{\frac{M-1}{2}}}{\Gam\left(\frac{M-1}{2}\right)}  h^{\frac{M-1}{2}} \int_{0}^1 (4th+1)^{\frac{1}{2}}(t)^{\frac{M-3}{2}}\,dt, \]
where the last equality has been obtained from the change of variable $x_m=ht$. 

Let us now compute the last integral which {only} converges when $M>1$. To that end, we first recall  Euler's integral representation formula for hypergeometric functions, see Theorem 2.2.1 \cite[p.~65]{MR1688958}. For $a,b,c\in \R$ such that $c>b>0$ , the hypergeometric function ${}_2F_1(a,b;c;z)$ can be written as
\begin{equation}\label{EulerHyp}
{}_2F_1(a,b;c;z)=\frac{\Gam(c)}{\Gam(b)\Gam(c-b)} \int_0^1 (1-zt)^{-a}\, t^{b-1} \,(1-t)^{c-b-1}\, dt.
\end{equation}
Writing $a=-\frac{1}{2}$, $b=\frac{M-1}{2}$, $c=b+1=\frac{M+1}{2}$ and $z=-4h$ we obtain for $M>1$ that
\[\int_{0}^1 (4th+1)^{\frac{1}{2}}(t)^{\frac{M-3}{2}}\,dt = \frac{2}{M-1} \; {}_2F_1\left(-\frac{1}{2},\frac{M-1}{2};\frac{M+1}{2};-4h\right),
\]
{from which the result follows.}
$\hfill\square$

\begin{remark}
Similar to the volume (see Remark \ref{CCVSP}), {by} (\ref{AreaSP}) {we obtain} an extension of the surface area of the corresponding paraboloids for the known cases $m=2$, $n=0$ and $m=3$, $n=0$. 
\begin{itemize}
\item In the case $m=2$, $n=0$, $SP_h^{2|0}$ is known to have the surface area (i.e. length in this case)  
\[\int_{-h^{\frac{1}{2}}}^{h^{\frac{1}{2}}} \left(1+4x^2_1\right)^{\frac{1}{2}} \, dx_1 = h^{\frac{1}{2}}\left(1+4h\right)^{\frac{1}{2}} + \frac{\sinh^{-1}(2h^{\frac{1}{2}})}{2} .\]
{Substituting} $M=2$ in (\ref{AreaSP}) we get {the same result:}
\[area(SP^{2|0}_h)= 2h^{\frac{1}{2}} {}_2F_1\left(-\frac{1}{2},\frac{1}{2};\frac{3}{2};-4h\right)=h^{\frac{1}{2}}\left(1+4h\right)^{\frac{1}{2}} + \frac{\sinh^{-1}(2h^{\frac{1}{2}})}{2}.\]
\item In the case $m=3$, $n=0$, the paraboloid $SP_h^{3|0}$ is known to have the surface area
\[2\pi \int_{0}^h x_3^{\frac{1}{2}} \left(1+\frac{1}{4x_3}\right)^{\frac{1}{2}} \,dx_3=\frac{\pi}{6} \left[(4h+1)^{\frac{3}{2}}-1\right].\]
{Substituting} $M=3$ in (\ref{AreaSP}) we {again get the same result:}
\[area(SP^{3|0}_h)=\pi h \; {}_2F_1\left(-\frac{1}{2},1;2;-4h\right)=\frac{\pi}{6} \left[(4h+1)^{\frac{3}{2}}-1\right].\]
\end{itemize}
\end{remark}

\subsection{Super-hyperboloid of revolution}
In 3 real dimensions, we consider the one-sheeted hyperboloid of revolution obtained by rotating the hyperbola $x_1^2-x_3^2=1$ around the $x_3$-axis. The Cartesian equation {of this} hyperboloid is 
\[x_1^2+x_2^2-x_3^2=1.\]
We define its extension to superspace by means of the superfunction 
\[g({\bf x})=\sum_{j=1}^{m-1} x_j^2 - x_m^2- \sum_{j=1}^n x\p_{2j-1}x\p_{2j}-1= -\hat{{\bf x}}^2-1-x_m^2. \hspace{.5cm} (m\geq 2).\]
Observe that the set {$\{\underline{x}\in\R^m: |\underline{\hat{x}}|^2-x_m^2-1\leq 0\}$ is non compact}. {However,} its intersection with the region $\{\underline{x}\in\R^m: x_m\in[-h,h] \}$ ($h>0$)  gives a compact set (symmetric with respect to the plane $x_m=0$) that is composed {of} the interior and the boundary of the hyperboloid $|\underline{\hat{x}}|^2-x_m^2=1$ in $\R^m$ with half height $h$. This means that we can integrate over the domain (and surface) defined by the superfunction $g$ with the restrictions $-h\leq x_m\leq h$.  This object will be called {the} {\it super-hyperboloid of revolution of half height $h>0$} and {is} denoted by $SH^{m|2n}_h$.  The domain and surface associated to $SH^{m|2n}_h$ are given by the characteristic functions
\[H(-g) H(h-x_m)H(h+x_m), \hspace{.5cm} \mbox{ {and} } \hspace{.5cm}\del(g)H(h-x_m)H(h+x_m),\]
respectively.

\begin{pro}\label{SH}
The volume of $SH^{m|2n}_h$ is {given by}
\begin{equation}\label{VolSH}
vol(SH^{m|2n}_h)= \begin{cases}\frac{2h\pi^{\frac{M-1}{2}}}{\Gam\left(\frac{M+1}{2}\right)}  \; {}_2F_1\left(\frac{1-M}{2}, \frac{1}{2}; \frac{3}{2}; -h^2\right), & M\notin -2\N+1,\\ 0, & M\in-2\N+1. \end{cases}
\end{equation}
\end{pro}
\pf 
Observe that 
\[vol(SH^{m|2n}_h)=\int_{\R^{m|2n}} H(-g) H(h-x_m) H(h+x_m)=\int_{-h}^h \left(\int_{\R^{m-1}}\int_B H(\hat{{\bf x}}^2+x_m^2+1)\, dV_{\underline{\hat{x}}}\right)dx_m,\]
where
 \[\int_B H(\hat{{\bf x}}^2+x_m^2+1)= \int_B H(x_m^2+1-|\underline{\hat{x}}|^2 + \underline{x}\p^2)= \sum_{j=0}^n \int_B \frac{\underline{x}\p^{\,2j}}{j!}\del^{(j-1)}(x_m^2+1-|\underline{\hat{x}}|^2)=\pi^{-n} \, \del^{(n-1)}(x_m^2+1-|\underline{\hat{x}}|^2).\]
Using (\ref{VolSB}) we obtain, 
\[\int_{\R^{m-1}}\int_B H(\hat{{\bf x}}^2+x_m^2+1)\, dV_{\underline{\hat{x}}}=\pi^{-n} \, \int_{\R^{m-1}} \del^{(n-1)}(x_m^2+1-|\underline{\hat{x}}|^2) \, dV_{\underline{\hat{x}}}=vol\left((x_m^2+1)^{\frac{1}{2}} \B^{m-1|2n}\right).\]
Hence, $\displaystyle\int_{\R^{m-1}}\int_B H(\hat{{\bf x}}^2+x_m^2+1)\, dV_{\underline{\hat{x}}}=\begin{cases}  \frac{\pi^{\frac{M-1}{2}}}{\Gam\left(\frac{M+1}{2}\right)} (x_m^2+1)^{\frac{M-1}{2}}, & M\notin -2\N+1,\\ 0, & M\in-2\N+1.\end{cases}$ This implies, for $M\in -2\N+1$, that $vol(SH^{m|2n}_h)=0$. But for $M\notin -2\N+1$  we have
\[vol(SH^{m|2n}_h)=\frac{2\pi^{\frac{M-1}{2}}}{\Gam\left(\frac{M+1}{2}\right)} \int_{0}^h (x_m^2+1)^{\frac{M-1}{2}}\, dx_m = \frac{h\pi^{\frac{M-1}{2}}}{\Gam\left(\frac{M+1}{2}\right)}  \int _0^1 (1+h^2t)^{\frac{M-1}{2}} t^{-\frac{1}{2}} \, dt,\]
where the last integral has been obtained from the change of variable $t=\frac{x_m^2}{h^2}$.
Using Euler's integral representation formula (\ref{EulerHyp}) for hypergeometric functions we get for $a=\frac{1-M}{2}$, $b=\frac{1}{2}$, $c=b+1=\frac{3}{2}$ and $z=-h^2$ that 
\[\frac{1}{2} \int _0^1 (1+h^2t)^{\frac{M-1}{2}} t^{-\frac{1}{2}} \, dt=  \; {}_2F_1\left(\frac{1-M}{2}, \frac{1}{2}; \frac{3}{2}; -h^2\right),\]
which completes the proof.
$\hfill\square$

\begin{remark}\label{RVSH}
Formula (\ref{VolSH}) constitutes an extension of the volume formulas for the corresponding hyperboloids in the classical cases $m=2$, $n=0$ and $m=3$, $n=0$. 
\begin{itemize}
\item In the case $m=2$, $n=0$, the hyperbola $SH_h^{2|0}=\{(x_1,x_2)\in\R^2: x_1^2-x_2^2\leq 1,\, -h\leq x_2\leq h\}$ is known to have the volume (i.e. area in $\R^2$)
\[2\int_{-h}^h (1+x_2^2)^{\frac{1}{2}} \, dx_2= 2\left[h(h^2+1)^{\frac{1}{2}}+\sinh^{-1}(h)\right],\]
{while evaluating} (\ref{VolSH}) for $M=2$ gives
\[vol(SH^{2|0}_h)=4h \;{}_2F_1\left(\frac{-1}{2}, \frac{1}{2}; \frac{3}{2}; -h^2\right)=2\left[h(h^2+1)^{\frac{1}{2}}+\sinh^{-1}(h)\right].\]
\item In the case $m=3$, $n=0$,  $SH_h^{3|0}=\{(x_1,x_2,x_3)\in\R^3: x_1^2+x_2^2-x_3\leq 1,\, -h\leq x_3\leq h\}$ is known to have the volume
\[\pi \int_{-h}^h (x_3^2+1)\,dx_3= 2\pi h\left(1+\frac{h^2}{3}\right),\]
{while} substituting $M=3$ in (\ref{VolSH}) {yields}
\[vol(SH^{3|0}_h)=2h\pi \; {}_2F_1\left(-1, \frac{1}{2}; \frac{3}{2}; -h^2\right)=2\pi h\left(1+\frac{h^2}{3}\right).\]
\end{itemize}
\end{remark}
\begin{pro}\label{AreaSH}
The surface area of $SH^{m|2n}_h$ is {given by}
\begin{equation}\label{AreaSH0}
area(SH^{m|2n}_h)=\begin{cases} \displaystyle \frac{4h\pi^{\frac{M-1}{2}}}{\Gam\left(\frac{M-1}{2}\right)} \, F_1\left(\frac{1}{2}; -\frac{1}{2}, \frac{3-M}{2}; \frac{3}{2}; -2h^2, -h^2\right), & M\notin -2\N+3,\\ 0, &  M\in -2\N+3,\end{cases}
\end{equation}
where $F_1(a; b_1, b_2; c; z_1, z_2)$ denotes Appell's hypergeometric function, see \cite[p.~73]{MR0185155}.
\end{pro}
\pf
Observe that 
\[area(SH^{m|2n}_h)= \int_{\R^{m|2n}} \del(g)\, |\pa_{\bf x}[g]| \, H(h-x_m) H(h+x_m)= \int_{-h}^h \left(\int_{\R^{m-1}} \int_B  \del(g)\, |\pa_{\bf x}[g]| \, dV_{\underline{\hat{x}}}\right)\, dx_m,\]
where $\pa_{\bf x}[g]=-(\pa_{\hat {\bf x}}-\pa_{x_m}e_m)(\hat{{\bf x}}^2+x_m^2+1)= -\pa_{\hat {\bf x}}[\hat{{\bf x}}^2]+\pa_{x_m}[x_m^2]e_m=-2\hat{{\bf x}}+2x_me_m$. Then, $|\pa_{\bf x}[g]|=2|\hat{{\bf x}}-x_me_m|= 2\left(-\hat{{\bf x}}^2+x_m^2\right)^{\frac{1}{2}}$. Using Proposition \ref{minus}, we write 
\[\del(g)\, |\pa_{\bf x}[g]| = 2\del\left(\hat{{\bf x}}^2+x_m^2+1\right)\left(-\hat{{\bf x}}^2+x_m^2\right)^{\frac{1}{2}}=K(-\hat{{\bf x}}^2),\]
where $K$ denotes the distribution $K(t)=2\del\left(x_m^2+1-t\right)\left(t+x_m^2\right)^{\frac{1}{2}}$. {Moreover} $$K(-\hat{{\bf x}}^2)=K(|\underline{\hat{x}}|^2-\underline{x}\p^{\,2})=\sum_{j=0}^n \frac{(-1)^n \underline{x}\p^{\,2j} }{j!} K^{(j)}(|\underline{\hat{x}}|^2),$$ {whence}
\[\int_B \del(g)\, |\pa_{\bf x}[g]|= \int_B K(-\hat{{\bf x}}^2)= (-1)^n \pi^{-n} K^{(n)}(|\underline{\hat{x}}|^2),\]
{and finally}
\[area(SH^{m|2n}_h)= \int_{-h}^h \left(\int_{\R^{m-1}} (-1)^n \pi^{-n} K^{(n)}(|\underline{\hat{x}}|^2) \, dV_{\underline{\hat{x}}} \right)\, dx_m.\]
Direct computations yield
\begin{align*}
(-1)^n \pi^{-n}  \int_{\R^{m-1}} K^{(n)}(|\underline{\hat{x}}|^2) \, dV_{\underline{\hat{x}}} &= (-1)^n \pi^{-n} \int_{\Sa^{m-2}} \left(\int_{0}^\infty  K^{(n)}(r^2) r^{m-2}\, dr \right) \, dS_{\underline{w}}\\
&= \frac{(-1)^n \pi^{-n} A_{m-1}}{2} \int_0^\infty K^{(n)}(t) t^{\frac{m-3}{2}} \, dt\\
&= \frac{ \pi^{-n} A_{m-1}}{2}  \int_0^\infty K(t) \frac{d^n}{dt^n}\left[t^{\frac{m-3}{2}}\right] \, dt   \\
&=  \frac{ \pi^{-n} A_{m-1}}{2} \prod_{j=1}^n \left(\frac{m-1}{2}-j\right) \int_0^\infty K(t) t^{\frac{M-3}{2}}\, dt,
\end{align*}
where for the special case $n=0$ we {put by convention} $\prod_{j=1}^0 \left(\frac{m-1}{2}-j\right):=1$. For  $\frac{M-3}{2}\in -\N$, it immediately follows that $ \prod_{j=1}^n \left(\frac{m-1}{2}-j\right)=0$ and in consequence $area(SH^{m|2n}_h)=0$. But for $\frac{M-3}{2}\notin -\N$ we have,
\begin{align}
(-1)^n \pi^{-n}  \int_{\R^{m-1}} K^{(n)}(|\underline{\hat{x}}|^2) \, dV_{\underline{\hat{x}}} &=
 \frac{ \pi^{-n} A_{m-1}}{2}  \frac{\Gam\left(\frac{m-1}{2}\right)}{\Gam\left(\frac{m-1}{2}-n\right)}\int_0^\infty 2\del\left(x_m^2+1-t\right)\left(t+x_m^2\right)^{\frac{1}{2}} t^{\frac{M-3}{2}}\, dt\nonumber\\
&= \frac{2\pi^{\frac{M-1}{2}}}{\Gam\left(\frac{M-1}{2}\right)} \left(2x_m^2+1\right)^{\frac{1}{2}} (x_m^2+1)^{\frac{M-3}{2}}.\nonumber
\end{align}
Thus,
\begin{align}
area(SH^{m|2n}_h) &=
\frac{4\pi^{\frac{M-1}{2}}}{\Gam\left(\frac{M-1}{2}\right)} \int_{0}^h \left(2x_m^2+1\right)^{\frac{1}{2}} (x_m^2+1)^{\frac{M-3}{2}} \, dx_m\nonumber\\
&= \frac{2h\pi^{\frac{M-1}{2}}}{\Gam\left(\frac{M-1}{2}\right)} \int_0^1(2h^2t+1)^{\frac{1}{2}} (h^2t+1)^{\frac{M-3}{2}} t^{-\frac{1}{2}} \, dt, \label{AS-HR1}
\end{align}
where the last equality has been obtained by the change of variable $t=\frac{x_m^2}{h^2}$.
The last integral can be written in terms of the so-called Appell's hypergeometric function of {the} first kind. Such a function constitutes an extension of the hypergeometric function of two variables and it is defined by 
\[F_1(a; b_1, b_2; c; z_1, z_2)=\sum_{j,k=0}^\infty \frac{(a)_{j+k}\,(b_1)_j\,(b_2)_k}{(c)_{j+k} \,j! \, k!} \,z_1^jz_2^k,\]
see \cite[Chapter~IX]{MR0185155} for more details. We {now} recall the integral representation of $F_1(a, b_1, b_2; c; z_1, z_2)$, see \cite[p.~77]{MR0185155}: 
\[F_1(a; b_1, b_2; c; z_1, z_2) = \frac{\Gam(c)}{\Gam(a)\Gam(c-a)}\int_0^1 t^{a-1} (1-t)^{c-a-1} (1-z_1t)^{-b_1} (1-z_2t)^{-b_2}\, dt, \hspace{.3cm} (c>a>0).\]
Taking $a=\frac{1}{2}$, $b_1=-\frac{1}{2}$, $b_2=\frac{3-M}{2}$, $c=a+1=\frac{3}{2}$, $z_1=-2h^2$ and $z_2=-h^2$ we obtain,
\begin{equation}\label{IntRepAH}
\int_0^1(2h^2t+1)^{\frac{1}{2}} (h^2t+1)^{\frac{M-3}{2}} t^{-\frac{1}{2}} \, dt= 2\, F_1\left(\frac{1}{2}; -\frac{1}{2}, \frac{3-M}{2}; \frac{3}{2}; -2h^2, -h^2\right).
\end{equation}
Finally, {substitution of} (\ref{IntRepAH}) into (\ref{AS-HR1}) {yields}  (\ref{AreaSH0}).
$\hfill\square$

\begin{remark}
Similarly to the previous results, (\ref{AreaSH0}) extends the known formulae for the surface area of the corresponding hyperboloids in the classical cases $m=2$, $n=0$ and $m=3$, $n=0$. 
\begin{itemize}
\item The hyperbola $SH_h^{2|0}$ (see Remark \ref{RVSH}) is known to have the surface area (i.e. length in this case)
\[S=4\int_1^{(1+h^2)^{\frac{1}{2}}} (2x_1^2-1)^{\frac{1}{2}} (x_1^2-1)^{-\frac{1}{2}} \, dx_1=2h\int_0^1 (2h^2t+1)^{\frac{1}{2}} t^{-\frac{1}{2}} (h^2t+1)^{-\frac{1}{2}}  \, dt, \]
where we have used the change of variable $x_1=(th^2+1)^{\frac{1}{2}}$, $0\leq t\leq 1$. {Then} (\ref{IntRepAH}) immediately shows that $S=4h\, F_1\left(\frac{1}{2}, -\frac{1}{2}, \frac{1}{2}; \frac{3}{2}; -2h^2, -h^2\right)$ which is the same result obtained when substituting $M=2$ in (\ref{AreaSH0}).
\item The hyperboloid $SH_h^{3|0}$ is known to have the surface area
\[2\pi \int_{-h}^h (2x_3^2+1)^{\frac{1}{2}}\, dx_3= \pi \left[2h(2h^2+1)^{\frac{1}{2}}+ 2^{\frac{1}{2}} \sinh^{-1}(2^{\frac{1}{2}}h)\right].\]
On the other hand, substituting $M=3$ in (\ref{AreaSH0}) we obtain
\[area(SH^{3|0}_h)=4h\pi \,  F_1\left(\frac{1}{2}; -\frac{1}{2}, 0; \frac{3}{2}; -2h^2, -h^2\right)=\pi \left[2h(2h^2+1)^{\frac{1}{2}}+ 2^{\frac{1}{2}} \sinh^{-1}(2^{\frac{1}{2}}h)\right].\]
 \end{itemize}
 \end{remark}


\section{Distributional Cauchy-Pompeiu formula in superspace}
The integration over general domains and surfaces in  superspace introduced in section \ref{Sec4} allows to extend and unify the {respective} known versions of the Cauchy-Pompeiu formulae in superspace.  In \cite{MR2539324, MR2521367}, the following Cauchy-Pompeiu formula was proven for bounded domains $\Om\inc \R^m$:
\begin{equation}\label{FlatC-PF}
\int_{\pa\Om}\int_B \nu_1^{m|2n}({\bf x}-{\bf y}) n(\underline{x}) G({\bf x}) \, dS_{\underline x}+\int_{\Om}\int_B \nu_1^{m|2n}({\bf x}-{\bf y}) \left(\pa_{\bf x}G({\bf x})\right) \, dV_{\underline x}=\begin{cases} - G({\bf y}), & \underline{y}\in {\Om},\\ 0, &  \underline{y}\notin {\ba{\Om}},\end{cases}
\end{equation}
where $\nu_1^{m|2n}$ is the fundamental solution of the super Dirac operator $\pa_{\bf x}$, see \cite{MR2386499}. The proof of this formula runs along similar lines {as} the traditional approach (see e.g.\ \cite[p.~147]{MR1169463}). It does not require the use of distributions  since it only considers integration over real sets composed with the Berezin integral. In \cite{MR2539324}, another version of the Cauchy-Pompeiu formula was obtained for the unit supersphere $\Sa^{m-1,2n}$. Its proof is based on (\ref{FlatC-PF}) and uses the distributional approach on the supersphere described in Example \ref{ExSS}:
\begin{equation}\label{SSC-PF}
\int_{\R^{m|2n}} \nu_1^{m|2n}({\bf x}-{\bf y}) 2\underline{x}\del({\bf x}^2+1) G({\bf x}) +\int_{\R^{m|2n}} \nu_1^{m|2n}({\bf x}-{\bf y}) H({\bf x}^2+1)  \left(\pa_{\bf x}G({\bf x})\right) =\begin{cases} -G({\bf y}), & \underline{y}\in {\B^{m}},\\ 0, &  \underline{y}\notin {\ba{\B^{m}}}.\end{cases}
\end{equation}
Both formulae generalize, in a certain way, the classical Cauchy-Pompeiu theorem in $\R^m$; see e.g.\ \cite[p.~147]{MR1169463}.  Yet this extension is not complete. Indeed, formulas (\ref{FlatC-PF}) and (\ref{SSC-PF}) only allow {for} integration over real domains (and surfaces) and the superball (and supersphere) respectively. In this section we show how our general integration approach allows to obtain a distributional Cauchy-Pompeiu formula in superspace for which (\ref{FlatC-PF})-(\ref{SSC-PF}) are obtained as particular cases.

We start with the following Stokes theorem in superspace.
\begin{teo}[{\bf Distributional Stokes Theorem, \cite{MR2539324}}]{}\label{StokTe}
Let $F,G\in C^\infty(\Om)\otimes \mathfrak{G}_{2n} \otimes \mathcal C_{m,2n}$ and $\al\in \mathcal{E}' \otimes \mathfrak{G}_{2n}^{(ev)}$ a distribution with compact support such that $supp\, \al\inc \Om\inc\R^m$. Then,
\begin{equation}\label{Stok}
\int_{\R^{m|2n}} \left(F\pa_{\bf x}\right)\al G+F\al \left(\pa_{\bf x}G\right)=-\int_{\R^{m|2n}} F\left(\pa_{\bf x}\al\right)G.
\end{equation}
\end{teo}
\pf
The proof is based in two fundamental {observations}: the support of $\al$ is compact and the operator $\int_B \pa_{x\p_j}$ is identically zero. Hence, for $F,G\in C^\infty(\Om)\otimes \mathfrak{G}_{2n}$ we have
\[\int_{\R^m}\pa_{x_j}\left[F\al G\right]=0, \hspace{1cm} \int_B\pa_{x\p_j}\left[F\al G\right]=0.\]
But $\pa_{\underline{x}} [F\al G]=(\pa_{\underline{x}}F)\al G+ F\al (\pa_{\underline{x}} G) + F(\pa_{\underline{x}} \al) G$, and by (\ref{Hash}) we get
\[\pa_{\underline{x}\p} [F^*\al G] = (\pa_{\underline{x}\p}F^*) \al G+ F(\pa_{\underline{x}\p}\, \al G)=  -(F\pa_{\underline{x}\p}) \al G+ F\al (\pa_{\underline{x}\p} G) + F(\pa_{\underline{x}\p} \, \al) G,\]
{whence}, 
\begin{align}
\int_{\R^{m|2n}}  (\pa_{\underline{x}}F)\al G+ F\al (\pa_{\underline{x}} G) &= -\int_{\R^{m|2n}}  F(\pa_{\underline{x}} \al) G,\label{S1}\\
\int_{\R^{m|2n}}   -(F\pa_{\underline{x}\p}) \al G+ F\al (\pa_{\underline{x}\p} G) &=  -\int_{\R^{m|2n}} F(\pa_{\underline{x}\p} \, \al) G. \label{S2}
\end{align}
Subtracting (\ref{S1}) from (\ref{S2}) we get (\ref{Stok}) for $F,G\in C^\infty(\Om)\otimes \mathfrak{G}_{2n}$. The extension to Clifford valued functions in $C^\infty(\Om)\otimes \mathfrak{G}_{2n} \otimes \mathcal C_{m,2n}$ is easily done by multiplying from the left and from the right {with} the corresponding Clifford generators $e_j, e_j\p$.
$\hfill\square$

In particular, if we take $\al=H(-g)$ with $g\in C^\infty(\R^m)\otimes \mathfrak{G}_{2n}^{(ev)}$ such that $\{g_0\leq 0\}$ is compact, we obtain a Stokes formula in superspace compatible with the notions of domain and surface integrals that we have introduced in section \ref{Sec4}.
\begin{cor}
Let $g=g_0+{\bf g}\in C^\infty(\R^m)\otimes \mathfrak{G}_{2n}^{(ev)}$ such that $\{g_0\leq 0\}$ is compact and $\pa_{\underline x}[g_0]\neq0$ on $g_0^{-1}(0)$. Then, for $F,G\in C^\infty(\Om)\otimes \mathfrak{G}_{2n} \otimes \mathcal C_{m,2n}$ such that $\{g_0\leq 0\}\inc \Om$ one has
\begin{equation}\label{StokPart}
\int_{\R^{m|2n}} H(-g )\left[\left(F\pa_{\bf x}\right) G+F \left(\pa_{\bf x}G\right)\right]= \int_{\R^{m|2n}} F\del(g) \pa_{\bf x}[g] G.
\end{equation}
\end{cor} 
\pf
The support of $H(-g)$ clearly {is} $\{g_0\leq 0\}$, which is compact. Then  (\ref{StokPart}) is the result of substituting $\al=H(-g)$ in (\ref{Stok}) and proving that $\pa_{\bf x}[H(-g)]=-\pa_{\bf x}[g]\del(g)$. Indeed, by (\ref{DervDelt}) we get
\begin{align*}
\pa_{x_k}[H(-g)] &= \pa_{x_k}[H(-g_0)]+\sum_{j=1}^n \pa_{x_k}\left[\frac{(-{\bf g})^j}{j!} \del^{(j-1)}(-g_0)\right] \\
&= -\del(g_0)\pa_{x_k}[g_0]+\sum_{j=1}^n (-1)^j  \left(\frac{({\bf g})^{j-1}}{(j-1)!} \pa_{x_k}[{\bf g}] \del^{(j-1)}(-g_0) - \frac{({\bf g})^{j}}{j!} \del^{(j)}(-g_0)\pa_{x_k}[g_0] \right)\\
&= -\pa_{x_k}[g_0] \sum_{j=0}^n \frac{(-{\bf g})^j}{j!} \del^{(j)}(-g_0) -  \pa_{x_k}[{\bf g}] \sum_{j=0}^{n-1} \frac{(-{\bf g})^j}{j!} \del^{(j)}(-g_0)\\
&= -\pa_{x_k}[g_0] \del(g) -  \pa_{x_k}[{\bf g}] \left(\del(g)-  \frac{{\bf g}^n}{n!} \del^{(n)}(g_0)\right)\\
&= -\pa_{x_k}[g] \del(g),
\end{align*}
{the} last equality {being obtained on account of} the nilpotency of the element $\pa_{x_k}[{\bf g}]$ which implies $\pa_{x_k}[{\bf g}]{\bf g}^n=0$. 

\noindent Using formula (\ref{Hash}) we easily get by induction that $\pa_{x\p_k}[{\bf g}^j]=j{\bf g}^{j-1}\pa_{x\p_k}[{\bf g}] $. Then,
\begin{align*}
\pa_{x\p_k}[H(-g)] &= \sum_{j=1}^n \pa_{x\p_k}\left[\frac{(-{\bf g})^j}{j!} \del^{(j-1)}(-g_0)\right]=  -  \sum_{j=1}^n  \frac{(-{\bf g})^{j-1}}{(j-1)!} \pa_{x\p_k}[{\bf g}] \del^{(j-1)}(-g_0)\\
&= -\pa_{x\p_k}[{\bf g}] \left(\del(g)-  \frac{{\bf g}^n}{n!} \del^{(n)}(g_0)\right)\\
&= -\pa_{x\p_k}[{\bf g}] \del(g).
\end{align*}
Now, one easily concludes that $\pa_{\bf x}[H(-g)]=-\pa_{\bf x}[g]\del(g)$. 
$\hfill\square$

In order to prove the distributional Cauchy-Pompeiu formula, we must observe first that we can substitute $F$ in the Stokes formula (\ref{Stok}) by any distribution with singular support disjoint from the singular support of $\al$. Let us make this more precise. We first recall that the singular support $sing \; supp \, \al$ of the distribution $\al\in\mathcal{D}'$ is defined by {the statement} that $\underline{x}\notin sing \; supp \, \al$ if and only if there exists a neighborhood $U_{\underline{x}}$ of $\underline{x}\in\R^m$ such that the restriction of $\al$ to $U_{\underline{x}}$ is a smooth function. Given two distributions $\al,\be\in \mathcal{D}'$ such that $sing \; supp \, \al \cap sing \; supp \, \be=\emptyset$, the product of distributions $\al\be$ is well defined {by} the formula
\begin{equation}\label{DistProd}
\langle \al\be, \phi\rangle= \langle \al, \be \chi \phi\rangle + \langle \be, \al (1-\chi) \phi\rangle, \hspace{1cm} \phi\in C^{\infty}(\R^m),
\end{equation}
where $\chi\in C^\infty(\R^m)$  is equal to zero in a neighborhood of $sing \; supp \, \be$ and equal to one in a neighborhood of $sing \; supp \, \al$.  It is easily seen that if $\al,\be\in \mathcal{D}'$ vanish in $\Om\inc\R^m$ then the product $\al\be$ vanishes in $\Om$ as well. Hence $supp\, \al\be \inc supp\,\al \cup supp\, \be$. As a consequence, if $\al$ and $\be$ have compact supports (i.e.\ $\al,\be\in  \mathcal{E}'$) then $\al\be$ also has compact support (i.e.\ $\al\be\in  \mathcal{E}'$). The product (\ref{DistProd}) is associative, commutative and satisfies the Leibniz rule. More information about multiplication of distributions can be found in \cite{MR3270560, MR1996773}. 

The notion of singular support can be extended to distributions $\al\in \mathcal{D}'\otimes \mathfrak{G}_{2n}$ by {the statement} that $\underline{x}\notin sing \; supp \, \al$ if and only if there exists a neighborhood $U_{\underline{x}}$ of $\underline{x}\in\R^m$ such that the restriction of $\al$  to $U_{\underline{x}}$ belongs to $C^\infty(U_{\underline x}) \otimes \mathfrak{G}_{2n}$.  {In} this way we obtain for every $\al\in\mathcal{D}'\otimes \mathfrak{G}_{2n}$ of the form (\ref{SupDistForm}) that
\[sing \; supp \, \al =\bigcup_{A\inc\{1,\ldots, n\}} sing \; supp \, \al_A. \]
In the same way, we define the product of superdistributions $\al, \be\in\mathcal{D}'\otimes \mathfrak{G}_{2n}$ with $sing \; supp \, \al \cap sing \; supp \, \be=\emptyset$ by
\begin{equation}\label{prodSupDist}
\al\be = \ \sum_{A,B\inc\{1,\ldots, n\}} \al_A\be_B\; \underline{x}\p_A\, \underline{x}\p_B,
\end{equation}
where the distribution $\al_A\be_B$ is {to be} understood in the sense of (\ref{DistProd}). 

Going back to the proof of the Stokes formula (\ref{Stok}), we have that
\[\int_{\R^m} \pa_{x_j}[\be\al G]\, dV_{\underline{x}}=0, \hspace{1cm} \mbox{ and } \hspace{1cm} \int_{B} \pa_{x\p_j}[\be\al G]=0.\]
where $\al \in \mathcal{E}'\otimes \mathfrak{G}_{2n}^{(ev)}$ and $\be\in \mathcal{E}'\otimes \mathfrak{G}_{2n}$  are such that $sing \; supp \, \al \cap sing \; supp \, \be=\emptyset$. Hence, we can repeat the proof of Theorem \ref{StokTe} to obtain (\ref{Stok}) but with $F\in \mathcal{E}'\otimes \mathfrak{G}_{2n}\otimes \mathcal{C}_{m,2n}$ with $sing \; supp \, \al \cap sing \; supp \, F=\emptyset$. Applying this reasoning to (\ref{StokPart}), for which we are considering $\al=H(-g)$, we immediately obtain the following consequence.

\begin{cor}
Let $g=g_0+{\bf g}\in C^\infty(\R^m)\otimes \mathfrak{G}_{2n}^{(ev)}$ such that $\{g_0\leq 0\}$ is compact and $\pa_{\underline x}[g_0]\neq0$ on $g_0^{-1}(0)$. Moreover let $\be\in \mathcal{E}'\otimes \mathfrak{G}_{2n}\otimes  \mathcal C_{m,2n}$ such that $sing \; supp \, \be \cap g^{-1}(0)=\emptyset$. Then, for every $G\in C^\infty(\Om)\otimes \mathfrak{G}_{2n} \otimes \mathcal C_{m,2n}$ such that $\{g_0\leq 0\}\inc \Om$ one has
\begin{equation}\label{StokPartD}
\int_{\R^{m|2n}} H(-g )\left[\left(\be\pa_{\bf x}\right) G+\be \left(\pa_{\bf x}G\right)\right]= \int_{\R^{m|2n}} \be\del(g) \pa_{\bf x}[g] G,
\end{equation}
where the distributional products $H(-g )(\be\pa_{\bf x})$, $H(-g )\be$, $\be\del(g)$ are understood in the sense of (\ref{prodSupDist})-(\ref{DistProd}).
\end{cor}   
\pf It suffices to note that $sing \; supp \, H(-g)= g_0^{-1}(0)$. $\hfill\square$

In \cite{MR2386499}, the fundamental solution of the super Dirac operator $\pa_{\bf x}$ was calculated to be,
\[\nu_1^{m|2n}=\pi^n \sum_{k=0}^{n-1} \frac{2^{2k+1} k!}{(n-k-1)!}\, \fhi_{2k+2}^{m|0} \underline{x}\p^{\, 2n-2k-1} - \pi^n \sum_{k=0}^n \frac{2^{2k} k!}{(n-k)!}\, \fhi_{2k+1}^{m|0} \underline{x}\p^{\, 2n-2k}, \]
where $ \fhi_j^{m|0}$ is the fundamental solution of $\pa_{\underline x}^j$. Observe that $\nu_1^{m|0}=-\fhi_1^{m|0}$. The superdistribution $\nu_1^{m|2n}$ satisfies 
\[\pa_{\bf x}\nu_1^{m|2n}({\bf x})=\del(\underline{x}) \frac{\pi^n}{n!} \underline{x}\p^{\, 2n} = \del({\bf x})= \nu_1^{m|2n}({\bf x}) \pa_{\bf x},\]
where $\del({\bf x})=\del(\underline{x}) \frac{\pi^n}{n!} \underline{x}\p^{\, 2n} $ defines the Dirac distribution on the supervector variable ${\bf x}$ and $\del(\underline{x})=\del(x_1)\cdots \del(x_m)$ is the $m$-dimensional real Dirac distribution. It is easily seen that
\begin{equation}\label{DelProp}
\langle \del({\bf x}-{\bf y}), G({\bf x})\rangle= \int_{\R^{m|2n}} \del({\bf x}-{\bf y})G({\bf x})=G({\bf y}) \mbox{ or equivalently, } \displaystyle\begin{cases} \displaystyle \int_{\R^m}  \del(\underline{x}- \underline{y}) G_A(\underline{x}) \, dV_{\underline{x}}=G_A(\underline{y}), \\ \\ \displaystyle\frac{\pi^n}{n!} \int_B (\underline{x}\p-\underline{y}\p)^{2n}\, \underline{x}\p_A= \underline{y}\p_A,\end{cases}
\end{equation}
where ${\bf y}=\underline{y}+\underline{y}\p$ and $G\in C^\infty(U_{\underline y})\otimes \mathfrak{G}_{2n}$ with $U_{\underline y}\inc \R^{m}$ being a neigborhood of ${\underline y}$.

In (\ref{StokPartD}) we can {effectuate} the substitution $\be=\nu_1^{m|2n}({\bf x}-{\bf y})$ with ${\bf y}=\underline{y}+\underline{y}\p$ such that $g_0(\underline{y})\neq 0$.  Indeed, it is easily seen that $sing \; supp \, \nu_1^{m|2n}({\bf x}-{\bf y})= \{\underline{y}\}$. {In} this way we get
\begin{multline}\label{IntB-PForm}
\int_{\R^{m|2n}} \del({\bf x}-{\bf y}) H(-g({\bf x})) G({\bf x}) \\
= \int_{\R^{m|2n}} \nu_1^{m|2n}({\bf x}-{\bf y})\, \del(g({\bf x}))\left(\pa_{{\bf x}}g({\bf x})\right) G({\bf x}) -   \int_{\R^{m|2n}} \nu_1^{m|2n}({\bf x}-{\bf y}) \, H(-g({\bf x})) \left(\pa_{\bf x}G({\bf x})\right).
\end{multline}
Let us now examine the distributional product 
\[\del({\bf x}-{\bf y}) H(-g({\bf x}))= \frac{\pi^n}{n!}(\underline{x}\p-\underline{y}\p)^{2n}  \sum_{j=0}^n  \frac{(-{\bf g}({\bf x}))^j}{j!} \; \del(\underline{x}-\underline{y}) \del^{(j-1)}(-g_0(\underline{x})).\]
It is clearly seen that $ sing \; supp \, \del(\underline{x}-\underline{y})=\{\underline{y}\}$ and  $ sing \; supp \, \del^{(j-1)}(-g_0(\underline{x}))=g_0^{-1}(0)$ {whence} (\ref{DistProd}) immediately shows for $g_0(\underline{y})\neq 0$ that 
\[\del(\underline{x}-\underline{y}) \del^{(j-1)}(-g_0(\underline{x}))=0, \hspace{1cm} j=1,\ldots, n.\]
Thus $\del({\bf x}-{\bf y}) H(-g({\bf x}))=\del({\bf x}-{\bf y}) H(-g_0({\underline x}))$ and then, (\ref{DelProp}) yields
\begin{equation}\label{FromDel-Heav}
\int_{\R^{m|2n}} \del({\bf x}-{\bf y}) H(-g({\bf x})) G({\bf x})=\int_{\R^{m|2n}} \del({\bf x}-{\bf y}) H(-g_0({\underline x})) G({\bf x})= H(-g_0({\underline y})) G({\bf y}).
\end{equation}
Substituting (\ref{FromDel-Heav}) into (\ref{IntB-PForm}) we obtain the following distributional Cauchy-Pompeiu formula in superspace.
\begin{teo}\label{C-P}
Let $g=g_0+{\bf g}\in C^\infty(\R^m)\otimes \mathfrak{G}_{2n}^{(ev)}$ such that $\{g_0\leq 0\}$ is compact and $\pa_{\underline x}[g_0]\neq0$ on $g_0^{-1}(0)$. Then, for $G\in C^\infty(\Om)\otimes \mathfrak{G}_{2n} \otimes \mathcal C_{m,2n}$ such that $\{g_0\leq 0\}\inc \Om$ one has
\begin{equation}\label{C-PForm}
\int_{\R^{m|2n}} \nu_1^{m|2n}({\bf x}-{\bf y})\, \del(g({\bf x})) \left(\pa_{{\bf x}}g({\bf x})\right) G({\bf x}) -   \int_{\R^{m|2n}} \nu_1^{m|2n}({\bf x}-{\bf y}) \, H(-g({\bf x})) \left(\pa_{\bf x}G({\bf x})\right)  = \begin{cases} G({\bf y}), & g_0(\underline{y})<0, \\ 0,  &  g_0(\underline{y})>0.  \end{cases}
\end{equation}
\end{teo}

As mentioned before, Theorem \ref{C-P} extends and unifies the known Cauchy-Pompeiu formulae (\ref{FlatC-PF}) and (\ref{SSC-PF}) (see {Theorems} 7 and 11 in \cite{MR2539324}). Indeed, in the particular case $g=g_0$, formula  (\ref{FlatC-PF}) immediately follows from (\ref{C-PForm}). On the other hand, for $g({\bf x})=-{\bf x}^2-1$, formula (\ref{C-PForm}) yields the supersphere case  (\ref{SSC-PF}).

\section*{Acknowledgements}
The authors want to thank Hendrik de Bie and Michael Wutzig for their careful reading of the manuscript and their very valuable suggestions. Alí Guzmán Adán is supported by a BOF-doctoral grant from Ghent University with grant number 01D06014.



\bibliographystyle{plain}

\begin{thebibliography}{10}

\bibitem{MR1688958}
George~E. Andrews, Richard Askey, and Ranjan Roy.
\newblock {\em Special functions}, volume~71 of {\em Encyclopedia of
  Mathematics and its Applications}.
\newblock Cambridge University Press, Cambridge, 1999.

\bibitem{MR0185155}
W.~N. Bailey.
\newblock {\em Generalized hypergeometric series}.
\newblock Cambridge Tracts in Mathematics and Mathematical Physics, No. 32.
  Stechert-Hafner, Inc., New York, 1964.

\bibitem{Berezin:1987:ISA:38130}
F.~A. Berezin.
\newblock {\em Introduction to Super Analysis}.
\newblock D. Reidel Publishing Co., Inc., New York, NY, USA, 1987.

\bibitem{1751-8121-42-24-245204}
H~De Bie, D~Eelbode, and F~Sommen.
\newblock Spherical harmonics and integration in superspace: Ii.
\newblock {\em Journal of Physics A: Mathematical and Theoretical},
  42(24):245204, 2009.

\bibitem{MR0208364}
Hans Bremermann.
\newblock {\em Distributions, complex variables, and {F}ourier transforms}.
\newblock Addison-Wesley Publishing Co., Inc., Reading, Mass.-London, 1965.

\bibitem{MR3270560}
Christian Brouder, Nguyen~Viet Dang, and Fr\'ed\'eric H\'elein.
\newblock A smooth introduction to the wavefront set.
\newblock {\em J. Phys. A}, 47(44):443001, 30, 2014.

\bibitem{MR2539324}
K.~Coulembier, H.~De~Bie, and F.~Sommen.
\newblock Integration in superspace using distribution theory.
\newblock {\em J. Phys. A}, 42(39):395206, 23, 2009.

\bibitem{MR2683546}
K.~Coulembier, H.~De~Bie, and F.~Sommen.
\newblock Orthosymplectically invariant functions in superspace.
\newblock {\em J. Math. Phys.}, 51(8):083504, 23, 2010.

\bibitem{de2007clifford}
H.~De~Bie and F.~Sommen.
\newblock A clifford analysis approach to superspace.
\newblock {\em Annals of Physics}, 322(12):2978--2993, 2007.

\bibitem{Bie2007}
H.~De~Bie and F.~Sommen.
\newblock Correct rules for clifford calculus on superspace.
\newblock {\em Advances in Applied Clifford Algebras}, 17(3):357--382, 2007.

\bibitem{MR2344451}
H.~De~Bie and F.~Sommen.
\newblock Spherical harmonics and integration in superspace.
\newblock {\em J. Phys. A}, 40(26):7193--7212, 2007.

\bibitem{MR2386499}
H.~De~Bie and F.~Sommen.
\newblock Fundamental solutions for the super {L}aplace and {D}irac operators
  and all their natural powers.
\newblock {\em J. Math. Anal. Appl.}, 338(2):1320--1328, 2008.

\bibitem{MR2521367}
H.~De~Bie and F.~Sommen.
\newblock A {C}auchy integral formula in superspace.
\newblock {\em Bull. Lond. Math. Soc.}, 41(4):709--722, 2009.

\bibitem{MR1169463}
R.~Delanghe, F.~Sommen, and V.~Sou{\v{c}}ek.
\newblock {\em Clifford algebra and spinor-valued functions}, volume~53 of {\em
  Mathematics and its Applications}.
\newblock Kluwer Academic Publishers Group, Dordrecht, 1992.
\newblock A function theory for the Dirac operator, Related REDUCE software by
  F. Brackx and D. Constales, With 1 IBM-PC floppy disk (3.5 inch).

\bibitem{MR778559}
Bryce DeWitt.
\newblock {\em Supermanifolds}.
\newblock Cambridge Monographs on Mathematical Physics. Cambridge University
  Press, Cambridge, 1984.

\bibitem{MR1996773}
Lars H\"ormander.
\newblock {\em The analysis of linear partial differential operators. {I}}.
\newblock Classics in Mathematics. Springer-Verlag, Berlin, 2003.
\newblock Distribution theory and Fourier analysis, Reprint of the second
  (1990) edition [Springer, Berlin; MR1065993 (91m:35001a)].

\bibitem{MR565567}
D.~A. Le\u\i~tes.
\newblock Introduction to the theory of supermanifolds.
\newblock {\em Uspekhi Mat. Nauk}, 35(1(211)):3--57, 255, 1980.

\bibitem{Pizz}
P.~Pizzetti.
\newblock Sulla media dei valori che una funzione dei punti dello spazio assume
  alla superficie di una sfera.
\newblock {\em Rend. Lincei}, (18):182--185, 1909.

\end{thebibliography}

\end{document}